\documentclass[
reprint,
dcolumn,
bm,
nofootinbib,
notitlepage]{revtex4-2}
\usepackage{graphicx}
\usepackage{color}
\usepackage[latin1]{inputenc}
\usepackage[T1]{fontenc}
\usepackage{amsmath}
\usepackage{amssymb}
\usepackage{braket}
\usepackage{mathrsfs}
\usepackage{hyperref}
\usepackage{ulem}
\usepackage{upgreek}
\usepackage{pstricks-add}
\usepackage{natbib} 
\usepackage{listings}
\usepackage{tikz}
\usepackage{tcolorbox}
\usetikzlibrary{mindmap}

\definecolor{lightblue}{rgb}{0.145,0.6666,1}

\begin{document}
\title{Beyond Cavity Born-Oppenheimer: On Non-Adiabatic Coupling and Effective Ground State Hamiltonians in Vibro-Polaritonic Chemistry}

\author{Eric W. Fischer}
\email{ericwfischer.sci@posteo.de}
\affiliation{Theoretische Chemie, Institut f\"ur Chemie, Universit\"at Potsdam,
Karl-Liebknecht-Stra\ss{}e 24-25, D-14476 Potsdam-Golm, Germany}

\author{Peter Saalfrank}
\affiliation{Theoretische Chemie, Institut f\"ur Chemie, Universit\"at Potsdam,
Karl-Liebknecht-Stra\ss{}e 24-25, D-14476 Potsdam-Golm, Germany}
\affiliation{Institut f\"ur Physik und Astronomie, Universit\"at Potsdam, Karl-Liebknecht-Stra\ss e 24-25, D-14476 Potsdam-Golm, Germany}

\date{\today}

\let\newpage\relax

\begin{abstract}
The emerging field of vibro-polaritonic chemistry studies the impact of light-matter hybrid states known as vibrational polaritons on chemical reactivity and molecular properties. Here, we discuss vibro-polaritonic chemistry from a quantum chemical perspective \textit{beyond} the cavity Born-Oppenheimer (CBO) approximation and examine the role of electron-photon correlation in effective ground state Hamiltonians. We first quantitatively review \textit{ab initio} vibro-polaritonic chemistry based on the molecular Pauli-Fierz Hamiltonian in dipole approximation and a vibrational strong coupling (VSC) Born-Huang expansion. We then derive non-adiabatic coupling elements arising from both ``slow'' nuclei and cavity modes compared to ``fast'' electrons via the generalized Hellmann-Feynman theorem, discuss their properties and re-evaluate the CBO approximation. In the second part, we introduce a \textit{crude} VSC Born-Huang expansion based on adiabatic electronic states, which provides a foundation for widely employed effective Pauli-Fierz Hamiltonians in ground state vibro-polaritonic chemistry. The latter \textit{do not} strictly respect the CBO approximation but an alternative scheme, which we name \textit{crude} CBO approximation. We argue that the \textit{crude} CBO ground state misses electron-photon entanglement relative to the CBO ground state due to neglected cavity-induced non-adiabatic transition dipole couplings to excited states. A perturbative connection between both ground state approximations is proposed, which identifies the \textit{crude} CBO ground state as first-order approximation to its CBO counterpart. We provide an illustrative numerical analysis of the cavity Shin-Metiu model with a focus on non-adiabatic coupling under VSC and electron-photon correlation effects on classical activation barriers. We finally discuss potential shortcomings of the electron-polariton Hamiltonian when employed in the VSC regime.
\end{abstract}

\let\newpage\relax
\maketitle
\newpage

\section{Introduction}
\label{sec.introduction}
Vibro-polaritonic chemistry is a rapidly evolving field at the intersection of chemistry and quantum optics, which studies the impact of vibrational polaritons on chemical reactivity and molecular properties.\cite{hirai2020,nagarajan2021,dunkelberger2022} Vibrational polaritons are light-matter hybrid states, which form when molecular vibrations interact strongly with quantized light modes of optical Fabry-P\'erot cavities operating in the infrared regime.\cite{ebbesen2016} The field of vibro-polaritonic chemistry was sparked by a series of seminal experiments, which reported on spectroscopic signatures of light-matter hybrid states\cite{shalabney2015,george2015,long2015,chervy2018,xiang2018,zhang2019} and, in particular, significantly altered reactive properties of chemical systems under vibrational strong coupling (VSC)\cite{george2016,thomas2016,thomas2019}. 

A quantum mechanical \textit{ab initio} description of molecular light-matter hybrid systems is based on the molecular Pauli-Fierz Hamiltonian underlying non-relativistic cavity quantum electrodynamics (cQED).\cite{craig1984,flick2017,ruggenthaler2022,flick2017cbo,schaefer2018,lihuo2021a} The Pauli-Fierz Hamiltonian accounts for interactions of electrons and nuclei forming molecules with quantized transverse field modes of an optical cavity and is usually formulated in dipole approximation and length-gauge representation. The presence of \textit{quantized} cavity modes renders the fully interacting light-matter hybrid scenario significantly more complex than the bare molecular many-body problem. Motivated by experimental findings\cite{thomas2016,thomas2019}, especially an electronic ground state theory for the VSC regime has been of significant interest in recent years. This stimulated the formulation of a cavity Born-Oppenheimer (CBO) approximation, which extends concepts from quantum chemistry to the realm of molecular cQED.\cite{flick2017,flick2017cbo} Within the CBO framework, both low-frequency cavity modes, \textit{e.g.}, infrared radiation, and nuclei are interpreted as ``slow'' degrees of freedom contrasted by ``fast'' electrons, which gives rise to two distinct types of non-adiabatic couplings.\cite{flick2017cbo} The CBO approximation manifests as truncation of an extended VSC Born-Huang expansion employing a basis of adiabatic electron-photon states, which is equivalent to neglecting both cavity and nuclear non-adiabatic couplings.\cite{flick2017cbo} A quantitative understanding and justification of the CBO approximation requires knowledge about the explicit nature of cavity and nuclear non-adiabatic coupling effects under VSC, which have only been discussed qualitatively so far.\cite{flick2017cbo} Motivated by those aspects, non-adiabatic coupling under VSC constitutes the first main topic of the present work. We explicitly derive analytic expressions for cavity and nuclear non-adiabatic derivative coupling elements from the generalized Hellmann-Feynman theorem\cite{gatti2017,cederbaum2004}, which allow us to access the detailed character of cavity and nuclear non-adiabatic effects under vibrational strong coupling.

In spirit of the CBO approximation, two distinct routes to ground state vibro-polaritonic chemistry emerged. The first one aims at solving an extended electronic structure problem, which is capable of accounting for correlations between electrons and cavity modes\cite{flick2017cbo},\textit{i.e.}, electron-photon correlation, by extending well established \textit{ab initio} methods from quantum chemistry to the realm of molecular cQED\cite{ruggenthaler2011,tokatly2013,ruggenthaler2014,flick2018,haugland2020,riso2022,bonini2022}. An alternative and especially popular approach relies in contrast on effective model Hamiltonians, which are obtained by projecting the molecular Pauli-Fierz Hamiltonian on the adiabatic electronic ground state\cite{delpino2015,galego2019,hernandez2019,angulo2020,subotnik2020,fischer2021,subotnik2021,li2021a,li2021b,yang2021,fischer2022a,fischer2022b,sun2022,lindoy2022a,lindoy2022b,mandal2022,
philbin2022,du2022,wang2022a,wang2022b,schaefer2022a,schaefer2022b,gomez2023,fischer2023,du2023} subsequently giving rise to even more approximate Tavis-Cummings\cite{tavis1968} or Dicke-models\cite{dicke1954} and variants thereof\cite{kockum2019}. Thus, ``only'' the correlated electronic ground state problem is solved \textit{ab initio}, which provides access to molecular potential energy and dipole moment surfaces by means of well established quantum chemistry methods. While the fully correlated approach to ground state vibro-polaritonic chemistry, here named VSC theory, has been consistently introduced in terms of an extended VSC Born-Huang expansion in combination with the CBO approximation\cite{flick2017,flick2017cbo}, effective approaches commonly lack a derivation and rely on \textit{ad hoc} formulations. 

The second aim of this paper is to close the gap between fully correlated and effective approaches to ground state vibro-polaritonic chemistry. For this purpose, we introduce a consistent formulation of effective approaches, denoted as \textit{crude} VSC theory, which employs a \textit{crude} VSC Born-Huang expansion in terms of adiabatic electronic states. We demonstrate the CBO approximation to be invalid with respect to effective ground state problems due to the emergence of non-adiabatic transition dipole couplings between adiabatic electronic states, which are not \textit{per se} small even for large energetic separation. In consequence, we formulate a \textit{crude} CBO approximation, which leads to widely employed effective model Hamiltonians in ground state vibro-polaritonic chemistry. We argue that the \textit{crude} CBO ground state misses electron-photon entanglement relative to the adiabatic electron-photon CBO ground state due to neglected non-adiabatic transition dipole couplings. 

In order to connect both ground state formulations, we propose a perturbative framework, which extends ideas from Herzberg-Teller theory to vibro-polaritonic chemistry. This new perspective allows us to perturbatively identify the \textit{crude} CBO ground state as first-order approximation to the adiabatic CBO ground state. Moreover, we discuss systematic corrections in the crude VSC framework, which account for electron-photon correlation. 

The paper is structured as follows: In Sec.\ref{sec.vsc_theory}, we provide a consistent quantitative discussion of \textit{ab initio} VSC theory beyond the cavity Born-Oppenheimer approximation and derive analytic expressions for nuclear and cavity non-adiabatic derivative couplings. In Sec.\ref{sec.crude_vsc_theory}, we introduce \textit{crude} VSC theory as foundation for effective ground state models in vibro-polaritonic chemistry based on a \textit{crude} CBO approximation. In Sec.\ref{sec.entanglement_perturbation}, we identify differences in electron-photon entanglement accounted for in approximate CBO and \textit{crude} CBO ground state theories and introduce a Herzberg-Teller inspired perturbative scheme for ground state VSC theory connecting both approaches. Sec.\ref{sec.numerical_analysis} numerically illustrates selected concepts via the cavity Shin-Metiu model. In Sec.\ref{sec.summary}, we summarize and conclude this paper.

\section{Vibrational Strong Coupling Theory for Molecules in Cavities}
\label{sec.vsc_theory}
We begin with a quantitative review of VSC theory based on the molecular Pauli-Fierz Hamiltonian in length-gauge representation and dipole approximation.\cite{flick2017,flick2017cbo} Starting from an extended VSC Born-Huang expansion, we explicitly derive coupled time-independent Schr\"odinger equations (TISE) for both adiabatic electron-photon states and nuclear-photon states along with non-adiabatic coupling elements for nuclei and cavity modes.

\subsection{The Molecular Pauli-Fierz Hamiltonian}
\label{subsec.velocity_pauli_fierz}
We consider a molecular system composed of $N_e$ electrons and $N_n$ nuclei coupled to $2N_c$ quantized transverse field modes of an infrared Fabry-P\'erot cavity. This light-matter hybrid system is described by the molecular Pauli-Fierz Hamiltonian in minimal-coupling form\cite{craig1984,flick2017,ruggenthaler2022}
\begin{multline}
\mathcal{\hat{H}}_\mathrm{PF}
=
\sum^{N_e}_i
\dfrac{\left(\underline{\hat{p}}_i+e\,\underline{\hat{A}}(\underline{r}_i)\right)^2}{2m_e}
+
\sum^{N_n}_a
\dfrac{\left(\underline{\hat{P}}_a-Q_a\,\underline{\hat{A}}(\underline{R}_a)\right)^2}{2M_a}
\\
+
V(\underline{r},\underline{R})
+
\sum^{2N_c}_{\lambda, k}
\hbar\omega_k
\left(
\hat{a}^\dagger_{\lambda k}
\hat{a}_{\lambda k}
+
\dfrac{1}{2}
\right)
\quad.
\label{eq.minimal_coupling_pauli_fierz}
\end{multline}
In the first line, we have kinetic energy contributions with electronic and nuclear momentum operators, $\underline{\hat{p}}_i=-\mathrm{i}\hbar\,\underline{\nabla}_i$, and, $\underline{\hat{P}}_a=-\mathrm{i}\hbar\,\underline{\nabla}_a$, electronic and nuclear masses, $m_e$ and $M_a$, elementary charge, $e$, as well as nuclear charges, $Q_a=Z_a e$, with charge number, $Z_a$. The first term in the second line is the molecular Coulomb-interaction potential 
\begin{multline}
V(\underline{r},\underline{R})
=
\underbrace{
\sum^{N_e}_{i>j}
\dfrac{e^2}{4\pi\epsilon_0\,\vert \underline{r}_i-\underline{r}_j\vert}}_{=V_{ee}(\underline{r})}
+
\underbrace{\sum^{N_n}_{a>b}
\dfrac{Q_a\,Q_b}{4\pi\epsilon_0\,\vert \underline{R}_a-\underline{R}_b\vert}}_{=V_{nn}(\underline{R})}
\\
\underbrace{-
\sum^{N_e}_i
\sum^{N_n}_a
\dfrac{Q_a\,e}{4\pi\epsilon_0\,\vert \underline{r}_i-\underline{R}_a\vert}}_{=V_{en}(\underline{r},\underline{R})}
\quad,
\label{eq.molecular_potential}
\end{multline}
with electron-electron, $V_{ee}(\underline{r})$, nuclear-nuclear, $V_{nn}(\underline{R})$, and electron-nuclear, $V_{en}(\underline{r},\underline{R})$, Coulomb interaction terms. Electronic, $\underline{r}_i$, and nuclear coordinates, $\underline{R}_a$, are collected in vectors, $\underline{r}=(\underline{r}_1,\dots,\underline{r}_{N_e})$, and, $\underline{R}=(\underline{R}_1,\dots,\underline{R}_{N_n})$. The last term in Eq.\eqref{eq.minimal_coupling_pauli_fierz} constitutes $2N_c$ quantized transverse field modes characterized by mode index, $k$, polarization index, $\lambda$, and harmonic mode frequency, $\omega_k$, respectively.\cite{cavity_mode_summation} Cavity modes are represented by photon creation and annihilation operators, $\hat{a}^\dagger_{\lambda k}$ and $\hat{a}_{\lambda k}$, which satisfy, $[\hat{a}_{\lambda k},\hat{a}^\dagger_{\lambda^\prime k^\prime}]=\delta_{\lambda\lambda^\prime}\delta_{kk^\prime}$. We restrict our discussion here to ideal cavities and neglect dissipative channels inducing spontaneous emission. 

The quantized vector potential, $\underline{\hat{A}}(\underline{r})$, in Eq.\eqref{eq.minimal_coupling_pauli_fierz} is purely transverse (Coulomb gauge), \textit{i.e.}, $\underline{\nabla}\cdot\underline{\hat{A}}(\underline{r})=0$, and given by\cite{craig1984}
\begin{align}
\underline{\hat{A}}(\underline{r})
&=
\sum^{2N_c}_{\lambda, k}
\dfrac{\underline{e}_{\lambda k}}{\omega_k}
\sqrt{\dfrac{\hbar\omega_k}{2\epsilon_0 V_c}}
\left(
\hat{a}^\dagger_{\lambda k}\,
e^{-\mathrm{i}\underline{k}_k\cdot\underline{r}}
+
\hat{a}_{\lambda k}\,
e^{\mathrm{i}\underline{k}_k\cdot\underline{r}}
\right)
\;,
\label{eq.vector_potential_sqr}
\vspace{0.2cm}
\\
&\approx 
\sum^{2N_c}_{\lambda, k}
\dfrac{\underline{e}_{\lambda k}}{\omega_k}
g_k
\left(
\hat{a}^\dagger_{\lambda k}
+
\hat{a}_{\lambda k}
\right)
\quad,
\label{eq.vector_potential_sqr_dipole_approximation}
\end{align}
with vacuum permittivity, $\epsilon_0$, and cavity quantization volume, $V_c$. Every cavity mode is doubly degenerate with respect to two orthogonal polarization directions related to polarization vectors, $\underline{e}_{\lambda k}$, which satisfy the orthonormality condition, $\underline{e}_{\lambda k}\cdot \underline{e}_{\lambda^\prime k^\prime}=\delta_{\lambda\lambda^\prime}\delta_{kk^\prime}$. In Eq.\eqref{eq.vector_potential_sqr_dipole_approximation}, we employed the dipole approximation, $e^{\pm\mathrm{i}\underline{k}_k\cdot\underline{r}}=1\pm\mathcal{O}(\mathrm{i}\,\underline{k}_k\cdot\underline{r})\approx 1$, which approximates, $\underline{\hat{A}}(\underline{r})\approx \underline{\hat{A}}$, as spatially uniform motivated by different length scales of infrared wavelengths and molecular diameters.\cite{schaefer2018} Further, we introduced a mode-specific light-matter interaction constant
\begin{align}
g_k
=
\sqrt{\frac{\hbar\omega_k}{2\epsilon_0 V_c}}
\quad.
\label{eq.light_matter_interaction_const}
\end{align}
From a quantum chemical perspective, the equivalent length-gauge representation of the molecular Pauli-Fierz Hamiltonian in dipole approximation is an appealing choice. It is obtained via\cite{craig1984,tokatly2013,mandal2020}
\begin{align}
\hat{\mathcal{U}}\,
\hat{\mathcal{S}}
&=
\exp
\biggl(
\dfrac{\mathrm{i}}{\hbar}
\left(
\underline{\hat{A}}
\cdot
\underline{d}_{en}
\right)
\biggr)
\exp
\biggl(
\mathrm{i}\,
\dfrac{\pi}{2}
\sum^{2N_c}_{\lambda, k}
\hat{n}_{\lambda k}
\biggr)
\quad,
\label{eq.length_gauge_transform}
\end{align}
where, $\hat{\mathcal{U}}$, mediates the unitary Power-Zienau-Woolley (PZW)\cite{power1959,woolley1971,babiker1974,woolley2020} transformation generated by, $\underline{\hat{A}}\cdot\underline{d}_{en}$, with molecular dipole moment
\begin{align}
\underline{d}_{en}
=
\underbrace{
-e
\sum^{N_e}_i
\underline{r}_i}_{\underline{d}_e}
+
\underbrace{
\sum^{N_n}_a
Q_a
\underline{R}_a}_{\underline{d}_n}
\quad,
\end{align}
exhibiting electronic, $\underline{d}_e$, and nuclear, $\underline{d}_n$, components. Further, $\hat{\mathcal{S}}$ in Eq.\eqref{eq.length_gauge_transform} relates to a unitary rotation in the cavity mode subspace, which is generated by the collective photon number operator, $\sum_{\lambda,k}\hat{n}_{\lambda k}=\sum_{\lambda,k}\hat{a}^\dagger_{\lambda k}\hat{a}_{\lambda k}$, and induces a real light-matter interaction term as discussed below.\cite{lihuo2021a} In length-gauge representation, the molecular Pauli-Fierz Hamiltonian in dipole approximation reads 
\begin{align}
\hat{H}_\mathrm{PF}
&=
\hat{\mathcal{S}}^\dagger\,
\hat{\mathcal{U}}^\dagger
\mathcal{\hat{H}}_\mathrm{PF}\,
\hat{\mathcal{U}}\,
\hat{\mathcal{S}}
\nonumber
\vspace{0.2cm}
\\
&=
\hat{T}_n
+
\hat{H}_e
+
\underbrace{\sum^{2N_c}_{\lambda, k}
\hbar\omega_k
\left(
\hat{a}^\dagger_{\lambda k}
\hat{a}_{\lambda k}
+
\dfrac{1}{2}
\right)}_{=\hat{H}_c}
\nonumber
\vspace{0.2cm}
\\
&
+
\underbrace{\sum^{2N_c}_{\lambda, k}
g_k\,
\biggl(
\underline{e}_{\lambda k}
\cdot
\underline{d}_{en}
\biggr)
\left(
\hat{a}^\dagger_{\lambda k}
+
\hat{a}_{\lambda k}
\right)}_{=\hat{H}_{sc}}
\nonumber
\vspace{0.2cm}
\\
&
+
\underbrace{\sum^{2N_c}_{\lambda, k}
\dfrac{g^2_k}{\hbar\omega_k}
\biggl(
\underline{e}_{\lambda k}
\cdot
\underline{d}_{en}
\biggr)^2}_{=\hat{H}_{dse}}
\quad.
\label{eq.length_pauli_fierz_sqr}
\end{align}
In the second line, we recover the nuclear kinetic energy operator (KEO), $\hat{T}_n$, the electronic Hamiltonian, $\hat{H}_e=\hat{T}_e+V(\underline{r},\underline{R})$, and the bare cavity Hamiltonian, $\hat{H}_c$. The third line constitutes the length-gauge representation of the light-matter interaction, $\hat{H}_{sc}$, determined by the polarization-projected molecular dipole moment, $\underline{e}_{\lambda k}\cdot\underline{d}_{en}$. Finally, the last term in Eq.\eqref{eq.length_pauli_fierz_sqr} resembles the dipole self-energy (DSE), $\hat{H}_{dse}$, which provides translational invariance, a bound ground state and gauge invariance of $\hat{H}_\mathrm{PF}$.\cite{rokaj2018,schaefer2020} 

In what follows, we aim at describing infrared cavity modes on the same footing as nuclear degrees of freedom, which is achieved via a cavity coordinate representation\cite{flick2017cbo}
\begin{align}
x_{\lambda k}
&=
\sqrt{\frac{\hbar}{2\omega_k}}
\left(
\hat{a}^\dagger_{\lambda k}
+
\hat{a}_{\lambda k}
\right)
\quad,
\label{eq.cavity_displacement_coord}
\vspace{0.2cm}
\\
\hat{p}_{\lambda k}
&=
\mathrm{i}
\sqrt{\frac{\hbar\omega_k}{2}}
\left(
\hat{a}^\dagger_{\lambda k}
-
\hat{a}_{\lambda k}
\right)
\quad.
\label{eq.cavity_momentum}
\end{align}  
Here, $x_{\lambda k}$ is a cavity displacement coordinate with conjugate momentum operator, $\hat{p}_{\lambda k}=-\mathrm{i}\hbar\frac{\partial}{\partial x_{\lambda k}}$, satisfying, $[x_{\lambda k},\hat{p}_{\lambda k}]=\mathrm{i}\hbar$. In coordinate presentation, the cavity Hamiltonian reads
\begin{align}
\hat{H}_c
&=
\underbrace{
\sum^{2N_c}_{\lambda, k}
\left(
\dfrac{\hat{p}^2_{\lambda k}}{2}
+
\dfrac{\omega^2_k}{2}
x^2_{\lambda k}
\right)}_{=\hat{T}_c
+
V_c(\underline{x})}
\quad,
\end{align}
with KEO, $\hat{T}_c$, and harmonic cavity potential, $V_c(\underline{x})$, where, $\underline{x}$, is a vector collecting all $2N_c$ displacement coordinates. The cavity coordinate representation of $\hat{H}_\mathrm{PF}$ is given by
\begin{multline}
\hat{H}_\mathrm{PF}
=
\hat{T}_n
+
\hat{T}_c
+
\hat{H}_e
\\
+
\underbrace{
\sum^{2N_c}_{\lambda, k}
\dfrac{\omega^2_k}{2}
\left(
x_{\lambda k}
+
\sqrt{\dfrac{2}{\hbar\omega^3_k}}\,
g_k
\biggl(
\underline{e}_{\lambda k}
\cdot
\underline{d}_{en}
\biggr)
\right)^2}_{=V_c+\hat{H}_{sc}+\hat{H}_{dse}}
\quad,
\label{eq.length_pauli_fierz_coord}
\end{multline}
which relates to a collection of displaced cavity modes shifted with respect to the polarization-projected molecular dipole moment, $\underline{e}_{\lambda k}\cdot\underline{d}_{en}$. One now introduces an electron-photon Hamiltonian\cite{flick2017cbo}
\begin{align}
\hat{H}_{ec}
&=
\hat{H}_e
+
V_c
+
\hat{H}_{sc}
+
\hat{H}_{dse}
\quad,
\label{eq.length_electron_photon_hamiltonian}
\end{align}
where the third term constitutes the length-gauge light-matter interaction in coordinate representation, 
\begin{align}
\hat{H}_{sc}
&=
\sum^{2N_c}_{\lambda, k}
\sqrt{\dfrac{2\omega_k}{\hbar}}\,g_k
\biggl(
\underline{e}_{\lambda k}
\cdot
\underline{d}_{en}
\biggr)
x_{\lambda k}
\quad.
\label{eq.length_light_matter_coupling}
\end{align}
The DSE term, $\hat{H}_{dse}$, remains as given in Eq.\eqref{eq.length_pauli_fierz_sqr}, since it is independent of the cavity mode representation. $\hat{H}_\mathrm{PF}$ in Eq.\eqref{eq.length_pauli_fierz_coord} can now be compactly rewritten as
\begin{align}
\hat{H}_\mathrm{PF}
&=
\hat{T}_n
+
\hat{T}_c
+
\hat{H}_{ec}
\quad,
\label{eq.length_pauli_fierz_short}
\end{align}
such that the corresponding light-matter TISE reads 
\begin{equation}
\left(
\hat{T}_n
+
\hat{T}_c
+
\hat{H}_{ec}
\right)
\ket{\Psi_i(\underline{R},\underline{x})}
=
\mathcal{E}_i
\ket{\Psi_i(\underline{R},\underline{x})}
\quad,
\label{eq.molecule_photon_tise_length}
\end{equation}
with energies, $\mathcal{E}_i$, and corresponding orthonormal light-matter many-body states, $\ket{\Psi_i(\underline{R},\underline{x})}$. The \textit{ket}-notation is employed with respect to the electronic subspace. In the remainder of this work, we are interested in finding approximate bound state solutions of Eq.\eqref{eq.molecule_photon_tise_length}.

\subsection{The VSC Born-Huang Expansion}
\label{subsec.length_vsc_expansion}
\textit{Ab initio} VSC theory is based on an extended Born-Huang expansion for light-matter many-body states, $\ket{\Psi_i(\underline{R},\underline{x})}$, in the following denoted as VSC Born-Huang expansion, given by\cite{flick2017cbo}
\begin{equation}
\ket{\Psi_i(\underline{R},\underline{x})}
=
\sum_{\mu}
\chi^{(nc)}_{i\mu}(\underline{R},\underline{x})\,
\ket{\Psi^{(ec)}_{\mu}(\underline{R},\underline{x})}
\quad.
\label{eq.vsc_born_huang}
\end{equation}
Here, $\ket{\Psi^{(ec)}_{\mu}(\underline{R},\underline{x})}$, are adiabatic electron-photon states, which parametrically depend on both nuclear and cavity coordinates, $(\underline{R},\underline{x})$, and provide a complete, orthonormal basis spanning the electronic subspace for each fixed configuration, $(\underline{R},\underline{x})$. Moreover, $\chi^{(nc)}_{i\mu}(\underline{R},\underline{x})$, are coefficient functions resembling nuclear-photon states. In order to indicate the hybrid character of states, we employ combined superscripts for electronic (e), nuclear (n) and cavity (c) contributions. We now insert Eq.\eqref{eq.vsc_born_huang} into the full light-matter TISE \eqref{eq.molecule_photon_tise_length} and project on adiabatic electron-photon states, $\bra{\Psi^{(ec)}_{\nu}}$, which leads to
\begin{align}
\sum_{\mu}
\bra{\Psi^{(ec)}_{\nu}}\,
\hat{H}_\mathrm{PF}
\left(
\chi^{(nc)}_{i\mu}
\ket{\Psi^{(ec)}_{\mu}}
\right)
&=
\mathcal{E}_i
\chi^{{(nc)}}_{i\nu}
\,.
\label{eq.projected_molecule_photon_tise_length}
\end{align}
$\ket{\Psi^{(ec)}_{\nu}}$ are taken to be eigenstates of $\hat{H}_{ec}$ in Eq.\eqref{eq.length_electron_photon_hamiltonian} and satisfy\cite{flick2017cbo}
\begin{align}
\hat{H}_{ec}\,
\ket{\Psi^{(ec)}_{\nu}(\underline{R},\underline{x})}
&=
E^{(ec)}_\nu(\underline{R},\underline{x})\,
\ket{\Psi^{(ec)}_{\nu}(\underline{R},\underline{x})}
\quad,
\label{eq.length_vsc_electron_photon_tise}
\end{align} 
with cavity potential energy surface (cPES), $E^{(ec)}_\nu(\underline{R},\underline{x})$, for the $\nu^\mathrm{th}$-state at fixed configuration, $(\underline{R},\underline{x})$. Further, KEOs, $\hat{T}_n$ and $\hat{T}_c$, in $\hat{H}_\mathrm{PF}$ act on both factors of the product, $\chi^{(nc)}_{i\mu}(\underline{R},\underline{x})\,\ket{\Psi^{(ec)}_{\mu}(\underline{R},\underline{x})}$, which leads to nuclear and cavity non-adiabatic coupling (NAC) elements between adiabatic electron-photon states\cite{fischer2022thesis}
\begin{align}
\mathcal{\hat{C}}^{(n)}_{\nu\mu}
&=
-
\sum^{N_n}_a
\dfrac{\hbar^2}{2M_a}
\left(
\mathcal{G}^{(n)}_{a,\nu\mu}
+
2\,\underline{\mathcal{F}}^{(n)}_{a,\nu\mu}
\cdot
\underline{\nabla}_a
\right)
\quad,
\label{eq.nuclear_derivative_nac}
\vspace{0.2cm}
\\
\mathcal{\hat{C}}^{(c)}_{\nu\mu}
&=
-
\dfrac{\hbar^2}{2}
\sum^{2N_c}_{\lambda,k}
\left(
\mathcal{G}^{(c)}_{\lambda k,\nu\mu}
+
2\,\mathcal{F}^{(c)}_{\lambda k,\nu\mu}
\dfrac{\partial}{\partial x_{\lambda k}}
\right)
\quad.
\label{eq.cavity_derivative_nac}
\end{align}  
Nuclear contributions are determined by 
\begin{align}
\begin{matrix}
\mathcal{G}^{(n)}_{a,\nu\mu}(\underline{R},\underline{x})
&=
\braket{
\Psi^{(ec)}_\nu(\underline{R},\underline{x})
\vert
\underline{\nabla}^2_a
\vert
\Psi^{(ec)}_\mu(\underline{R},\underline{x})}_{\underline{r}}
\quad,
\vspace{0.2cm}
\\
\underline{\mathcal{F}}^{(n)}_{a,\nu\mu}(\underline{R},\underline{x})
&=
\braket{
\Psi^{(ec)}_\nu(\underline{R},\underline{x})
\vert
\underline{\nabla}_a
\vert
\Psi^{(ec)}_\mu(\underline{R},\underline{x})}_{\underline{r}}
\quad,
\end{matrix}
\label{eq.nuclear_derivative_nac_gf}
\end{align}
and cavity contributions decompose into\cite{fischer2022thesis,schnappinger2023}
\begin{align}
\begin{matrix}
\mathcal{G}^{(c)}_{\lambda k,\nu\mu}(\underline{R},\underline{x})
&=
\braket{
\Psi^{(ec)}_\nu(\underline{R},\underline{x})
\vert
\dfrac{\partial^2}{\partial x^2_{\lambda k}}
\vert
\Psi^{(ec)}_\mu(\underline{R},\underline{x})}_{\underline{r}}
\,,
\vspace{0.2cm}
\\
\mathcal{F}^{(c)}_{\lambda k,\nu\mu}(\underline{R},\underline{x})
&=
\braket{
\Psi^{(ec)}_\nu(\underline{R},\underline{x})
\vert
\dfrac{\partial}{\partial x_{\lambda k}}
\vert
\Psi^{(ec)}_\mu(\underline{R},\underline{x})}_{\underline{r}}
\,,
\end{matrix}
\label{eq.cavity_derivative_nac_gf}
\end{align}
where integration with respect to electronic coordinates is indicated by $\braket{\dots}_{\underline{r}}$. Note, both cavity NAC elements are scalar in nature, in contrast to the first-order nuclear derivative term, $\underline{\mathcal{F}}^{(n)}_{a,\nu\mu}$, which is a vector. The length-gauge nuclear-photon TISE is eventually obtained as
\begin{multline}
\left(
\hat{T}_n
+
\hat{T}_c
+
E^{(ec)}_\nu(\underline{R},\underline{x})
\right)
\chi^{(nc)}_{i\nu}
\\
+
\sum_{\mu\neq\nu}
\mathcal{\hat{K}}_{\nu\mu}\,
\chi^{(nc)}_{i\mu}
=
\mathcal{E}_i\,
\chi^{(nc)}_{i\nu}
\quad,
\label{eq.vsc_length_nuclear_photon_tise}
\end{multline}
with generalized NAC elements given by
\begin{align}
\mathcal{\hat{K}}_{\nu\mu}
&=
\mathcal{\hat{C}}^{(n)}_{\nu\mu}
+
\mathcal{\hat{C}}^{(c)}_{\nu\mu}
\quad.
\label{eq.length_vsc_general_nacs}
\end{align}
In the following, we give a detailed characterization of $\mathcal{\hat{K}}_{\nu\mu}$, which lies at the heart of the cavity Born-Oppenheimer (CBO) approximation.  

\subsection{Non-Adiabatic Coupling under VSC and the CBO Approximation}
\label{subsec.length_nacs_details}
We evaluate first-order derivative NAC elements, $\mathcal{F}^{(c)}_{\lambda k,\nu\mu}$ and $\underline{\mathcal{F}}^{(n)}_{a,\nu\mu}$, analytically by means of the generalized (alternatively off-diagonal) Hellmann-Feynman theorem\cite{gatti2017,cederbaum2004}
\begin{align}
\mathcal{F}^{(c)}_{\lambda k,\nu\mu}
&=
\dfrac{\braket{\Psi^{(ec)}_\nu
\vert
\dfrac{\partial}{\partial x_{\lambda k}} 
\hat{H}_{ec}
\vert
\Psi^{(ec)}_\mu}_{\underline{r}}}{E^{(ec)}_\mu-E^{(ec)}_\nu}
\quad,
\label{eq.hellmann_feynman_cavity}
\vspace{0.2cm}
\\
\underline{\mathcal{F}}^{(n)}_{a,\nu\mu}
&=
\dfrac{\braket{\Psi^{(ec)}_\nu
\vert
\underline{\nabla}_a
\hat{H}_{ec}
\vert
\Psi^{(ec)}_\mu}_{\underline{r}}}{E^{(ec)}_\mu-E^{(ec)}_\nu}
\quad,
\label{eq.hellmann_feynman_nuclei}
\end{align}
where, $\hat{H}_{ec}$, is the electron-photon Hamiltonian in Eq.\eqref{eq.length_electron_photon_hamiltonian}. For the cavity contribution, we obtain (\textit{cf.} Appendix \ref{subsec.appendix_derivative_nacs} for details)
\begin{align}
\mathcal{F}^{(c)}_{\lambda k,\nu\mu}
&=
\sqrt{\dfrac{2\omega_k}{\hbar}}\,
g_k\,
\dfrac{
\biggl(
\underline{e}_{\lambda k}
\cdot
\underline{\mathcal{D}}_{\nu\mu}
\biggr)}{E^{(ec)}_\mu-E^{(ec)}_\nu}
\quad,
\label{eq.length_cavity_nac_hellmann_feynman}
\end{align}
with transition dipole elements between adiabatic electron-photon states
\begin{align}
\underline{\mathcal{D}}_{\nu\mu}(\underline{R},\underline{x})
=
\braket{
\Psi^{(ec)}_\nu(\underline{R},\underline{x})
\vert
\underline{d}_{en}
\vert
\Psi^{(ec)}_\mu(\underline{R},\underline{x})}_{\underline{r}}
\quad.
\end{align} 
Eq.\eqref{eq.length_cavity_nac_hellmann_feynman} can be traced back to contributions from the light-matter interaction term, $\hat{H}_{sc}$. Further, since $\mathcal{F}^{(c)}_{\lambda k,\nu\mu}$ is directly proportional to polarization-projected transition dipole moments, $\underline{e}_{\lambda k}\cdot\underline{\mathcal{D}}_{\nu\mu}$, the optical character of adiabatic electron-photon states plays a central role: Cavity non-adiabatic coupling elements vanish for dark transitions characterized by, $\underline{\mathcal{D}}_{\nu\mu}=0$, and are non-zero for bright transitions with $\underline{\mathcal{D}}_{\nu\mu}\neq0$, where, $\mathcal{F}^{(c)}_{\lambda k,\nu\mu}$, becomes particularly large for energetically close-lying adiabatic states.

For nuclear derivative NAC elements in Eq.\eqref{eq.hellmann_feynman_nuclei}, we find (\textit{cf.} Appendix \ref{subsec.appendix_derivative_nacs})
\begin{multline}
\underline{\mathcal{F}}^{(n)}_{a,\nu\mu}
=
\dfrac{\braket{\Psi^{(ec)}_\nu
\vert
\underline{\nabla}_a
\hat{V}_{en}
\vert
\Psi^{(ec)}_\mu}_{\underline{r}}}
{E^{(ec)}_\mu-E^{(ec)}_\nu}
\\
+
\displaystyle\sum^{2N_c}_{\lambda, k}
\dfrac{2\,Q_a\,g^2_k}{\hbar\omega_k}
\dfrac{
\biggl(
\underline{e}_{\lambda k}
\cdot
\underline{\mathcal{D}}_{\nu\mu}
\biggr)\,
\underline{e}_{\lambda k}}
{E^{(ec)}_\mu-E^{(ec)}_\nu}
\quad,
\label{eq.length_nuclear_nac_hellmann_feynman}
\end{multline}
where the first term resembles the vibronic non-adiabatic coupling in the subspace of adiabatic \textit{electron-photon} states. The second term constitutes a cavity-induced matter contribution, which emerges from the dipole self-energy and can be related to the transverse polarization operator (\textit{cf.} Appendix \ref{subsec.appendix_derivative_nacs}). Thus, \textit{bright} and \textit{dark} adiabatic electron-photon states do not only behave differently with respect to cavity-induced non-adiabatic coupling but also with respect to nuclear contributions under VSC. 

When the adiabatic electron-photon ground state is energetically well separated from the excited state manifold, \textit{i.e.}, $\vert E^{(ec)}_\mu-E^{(ec)}_0\vert\gg0$, both cavity and nuclear NAC elements are small and the VSC Born-Huang expansion can be truncated to the first term\cite{flick2017cbo}
\begin{align}
\ket{\Psi_\mathrm{cbo}(\underline{R},\underline{x})}
&=
\chi^{(nc)}_{0}(\underline{R},\underline{x})\,
\ket{\Psi^{(ec)}_{0}(\underline{R},\underline{x})}
\quad.
\label{eq.vsc_cbo_ground_state}
\end{align}
This approach resembles the CBO approximation to the light-matter many-body ground state, $\ket{\Psi_0(\underline{R},\underline{x})}\approx\ket{\Psi_\mathrm{cbo}(\underline{R},\underline{x})}$, where the corresponding electron-photon ground state satisfies
\begin{align} 
\hat{H}_{ec}
\ket{\Psi^{(ec)}_0(\underline{R},\underline{x})}
&=
\mathcal{E}^{(ec)}_0(\underline{R},\underline{x})\,
\ket{\Psi^{(ec)}_0(\underline{R},\underline{x})}
\quad.
\label{eq.vsc_ground_state_tise}
\end{align}
Two approximate schemes concerning NAC elements are now possible: The first one, here denoted as \textit{cavity adiabatic} approximation in analogy to vibronic notation\cite{ballhausen1972}, neglects off-diagonal NAC elements and accounts for diagonal CBO ground state corrections only. The corresponding approximate adiabatic ground state cPES reads  
\begin{align}
\mathcal{E}^{(ec)}_0(\underline{R},\underline{x})
&=
E^{(ec)}_0(\underline{R},\underline{x})
+
\mathcal{G}^{(n)}_{00}
+
\mathcal{G}^{(c)}_{00}
\quad,
\label{eq.adiabatic_vsc_approximation}
\end{align}
with
\begin{align}
\mathcal{G}^{(n)}_{00}
=
-
\sum^{N_n}_a
\dfrac{\hbar^2}{2M_a}
\mathcal{G}^{(n)}_{a,00}
\hspace{0.2cm}
,
\hspace{0.2cm}
\mathcal{G}^{(c)}_{00}
=
-
\dfrac{\hbar^2}{2}
\sum^{2N_c}_{\lambda,k}
\mathcal{G}^{(c)}_{\lambda k,00}
\;.
\end{align}
Here, $\mathcal{G}^{(n)}_{a,00}$ and $\mathcal{G}^{(c)}_{\lambda k,00}$ are defined in Eqs.\eqref{eq.nuclear_derivative_nac_gf} and \eqref{eq.cavity_derivative_nac_gf}. The second scheme is realized by additionally setting diagonal non-adiabatic contributions to zero, \textit{i.e.}, $\mathcal{\hat{K}}_{\nu\mu}=0$, for all $\nu,\mu$, such that, $\mathcal{E}^{(ec)}_0(\underline{R},\underline{x})\approx E^{(ec)}_0(\underline{R},\underline{x})$, in Eq.\eqref{eq.vsc_ground_state_tise} is the bare cPES. Note, our definition slightly differs from the original formulation in Ref.\cite{flick2017cbo}, where the CBO approximation took into account diagonal NAC contributions, $\mathcal{G}^{(n)}_{00}$, and, $\mathcal{G}^{(c)}_{00}$.

\section{Crude VSC Theory for Molecules in Cavities}
\label{sec.crude_vsc_theory}
\subsection{The Crude VSC Born-Huang Expansion}
\label{subsec.length_crude_vsc_expansion}
VSC theory, as discussed in the last section, crucially relies on the knowledge of adiabatic electron-photon states obtained from TISE \eqref{eq.length_vsc_electron_photon_tise}. We now introduce an alternative approach, denoted as \textit{crude} VSC theory, which circumvents this issue by employing a basis of adiabatic electronic states. In crude VSC theory, we consider a \textit{crude} VSC Born-Huang expansion
\begin{equation}
\ket{\Psi_i(\underline{R},\underline{x})}
=
\sum_{\mu}
\tilde{\chi}^{(nc)}_{i\mu}(\underline{R},\underline{x})\,
\ket{\Psi^{(ec)}_{\mu}(\underline{R},\underline{x}_0)}
\quad,
\label{eq.crude_vsc_born_huang_prelim}
\end{equation} 
where adiabatic electron-photon states, $\ket{\Psi^{(ec)}_{\mu}(\underline{R},\underline{x}_0)}$, are considered at a fixed cavity displacement coordinate reference value, $\underline{x}_0$. In order to make the difference to VSC theory explicit, we write nuclear-cavity states, $\tilde{\chi}^{(nc)}_{i\mu}(\underline{R},\underline{x})$, augmented by a tilde sign. We now choose reference cavity displacement coordinates as
\begin{align}
x_{0,\lambda k}
=
-
\sqrt{\dfrac{2}{\hbar\omega^3_k}}\,g_k
\biggl(
\underline{e}_{\lambda k}
\cdot
\underline{d}_{en}
\biggr)
\quad,
\label{eq.length_cavity_reference_config}
\end{align}
such that the electron-photon Hamiltonian reduces to the bare electronic Hamiltonian, $\hat{H}_{ec}(\underline{x}_0)=\hat{H}_e$, for the cavity reference configuration, $\underline{x}_0$. Hence, the electron-photon TISE \eqref{eq.length_vsc_electron_photon_tise} evaluated at $\underline{x}_0$ reduces to the bare electronic TISE
\begin{equation}
\hat{H}_e
\ket{\Psi^{(e)}_{\nu}(\underline{R})}
=
E^{(e)}_\nu(\underline{R})\,
\ket{\Psi^{(e)}_{\nu}(\underline{R})}
\quad,
\label{eq.vsc_zero_electron_photon_tise}
\end{equation}
which is solved by adiabatic \textit{electronic} states, \textit{i.e.}, $\ket{\Psi^{(ec)}_{\nu}(\underline{R},\underline{x}_0)}=\ket{\Psi^{(e)}_{\nu}(\underline{R})}$, with molecular potential energy surfaces, $E^{(e)}_\nu(\underline{R})$. Accordingly, the crude VSC expansion, Eq.\eqref{eq.crude_vsc_born_huang_prelim}, in combination with Eq.\eqref{eq.length_cavity_reference_config} reduces to
\begin{equation}
\ket{\Psi_i(\underline{R},\underline{x})}
=
\sum_{\mu}
\tilde{\chi}^{(nc)}_{i\mu}(\underline{R},\underline{x})\,
\ket{\Psi^{(e)}_{\mu}(\underline{R})}
\quad,
\label{eq.crude_vsc_born_huang}
\end{equation}
where the adiabatic basis is spanned by conventional adiabatic \textit{electronic} states, which only exhibit a parametric dependence on nuclear coordinates.

\subsection{Coupled TISE in Crude VSC Theory}
\label{subsec.coupled_tise_vsc_theory}
We now derive coupled TISE in crude VSC theory by inserting Eq.\eqref{eq.crude_vsc_born_huang} into the light-matter TISE \eqref{eq.molecule_photon_tise_length}, which leads after projection on $\bra{\Psi^{(e)}_{\nu}}$ to
\begin{align}
\sum_{\mu}
\bra{\Psi^{(e)}_{\nu}}\,
\hat{H}_\mathrm{PF}
\left(
\tilde{\chi}^{(nc)}_{i\mu}\,
\ket{\Psi^{(e)}_{\mu}}
\right)
=
\mathcal{E}_i\,
\tilde{\chi}^{{(nc)}}_{i\nu}
\quad.
\label{eq.crude_projected_molecule_photon_tise_length}
\end{align}
Evaluation of matrix elements, as shown in Appendix \ref{subsec.crude_vsc_details}, eventually provides the bare electronic TISE \eqref{eq.vsc_zero_electron_photon_tise} and the crude nuclear-photon TISE 
\begin{multline}
\left(
\hat{T}_n
+
\hat{T}_c
+
V_\nu(\underline{R},\underline{x})
\right)
\tilde{\chi}^{(nc)}_{i\nu}
\\
+
\sum_{\mu\neq\nu}
\hat{K}_{\nu\mu}\,
\tilde{\chi}^{(nc)}_{i\mu}
=
\mathcal{E}_i\,
\tilde{\chi}^{(nc)}_{i\nu}
\quad,
\label{eq.vsc_zero_length_nuclear_photon_tise}
\end{multline}
with \textit{crude} cPES
\begin{multline}
V_\nu(\underline{R},\underline{x})
=
E^{(e)}_\nu(\underline{R})
+
V_c(\underline{x})
\\
+
\sum^{2N_c}_{\lambda,k}
\sqrt{\dfrac{2\omega_k}{\hbar}}\,
g_k
\biggl(
\underline{e}_{\lambda k}
\cdot
\underline{d}_{\nu\nu}
\biggr)\,
x_{\lambda k}
\\
+
\sum^{2N_c}_{\lambda,k}
\dfrac{g^2_k}{\hbar\omega_k}
\biggl[
\biggl(
\underline{e}_{\lambda k}
\cdot
\underline{d}^2_{\nu\nu}
\biggr)
+
\sum_{\alpha\neq\nu}
\biggl(
\underline{e}_{\lambda k}
\cdot
\underline{d}^2_{\nu\alpha}
\biggr)
\biggr]
\quad,
\label{eq.crude_vsc_length_cpes}
\end{multline}
and dipole matrix elements
\begin{align}
\underline{d}_{\nu\mu}
&=
\braket{
\Psi^{(e)}_{\nu}
\vert
\underline{d}_{en}
\vert
\Psi^{(e)}_{\mu}
}_{\underline{r}}
\quad.
\label{eq.crude_dipole_matrix_elements}
\end{align}
Eventually, generalized NAC elements read
\begin{multline}
\hat{K}_{\nu\mu}
=
\hat{C}^{(n)}_{\nu\mu}
+
\sum^{2N_c}_{\lambda,k}
\sqrt{\dfrac{2\omega_k}{\hbar}}\,
g_k
\biggl(
\underline{e}_{\lambda k}
\cdot
\underline{d}_{\nu\mu}
\biggr)\,
x_{\lambda k}
\\
+
\sum^{2N_c}_{\lambda,k}
\dfrac{g^2_k}{\hbar\omega_k}
\sum_\alpha
\biggl(
\underline{e}_{\lambda k}
\cdot
\underline{d}_{\nu\alpha}
\biggr)
\biggl(
\underline{e}_{\lambda k}
\cdot
\underline{d}_{\alpha\mu}
\biggr)
\quad,
\label{eq.crude_length_vsc_general_nacs}
\end{multline}
where the bare nuclear derivative coupling element, $\hat{C}^{(n)}_{\nu\mu}$, is augmented by two non-adiabatic coupling terms related to polarization-projected transition dipole moments, $\underline{e}_{\lambda k}\cdot\underline{d}_{\nu\mu}$. Those terms emerge as off-diagonal contributions of light-matter interaction and DSE terms in the adiabatic electronic basis.

\subsection{Crude Non-Adiabatic Coupling and the Crude CBO Approximation}
\label{subsec.crude_cbo_approx}
We shall now examine the detailed character of crude non-adiabatic coupling elements, $\hat{K}_{\nu\mu}$, introduced in Eq.\eqref{eq.crude_length_vsc_general_nacs} and the notion of a CBO approximation in \textit{crude} VSC theory. We first recall, that nuclear derivative NAC elements follow from the generalized Hellmann-Feynman theorem as\cite{cederbaum2004}
\begin{align}
\underline{F}^{(n)}_{a,\nu\mu}
&=
\dfrac{\braket{\Psi^{(e)}_\nu
\vert
\underline{\nabla}_a\hat{H}_{e}
\vert
\Psi^{(e)}_\mu}_{\underline{r}}}{E^{(e)}_\mu-E^{(e)}_\nu}
\quad,
\label{eq.nuclear_nac_hellmann_feynman}
\end{align}
which become small for energetically well separated states. However, for cavity-induced NAC contributions in Eq.\eqref{eq.crude_length_vsc_general_nacs}, we found expressions crucially different from Eq.\eqref{eq.nuclear_nac_hellmann_feynman}: Those terms are proportional to transition dipole moments, $\underline{d}_{\nu\mu}$, which are in general not necessarily small for energetically well separated states and therefore violate the condition for invoking the CBO approximation.

In order to make progress, we introduce a truncation of the crude VSC expansion, Eq.\eqref{eq.crude_vsc_born_huang}, in analogy to Eq.\eqref{eq.vsc_cbo_ground_state} as
\begin{align}
\ket{\Psi_\mathrm{ccbo}(\underline{R},\underline{x})}
&=
\tilde{\chi}^{(nc)}_{0}(\underline{R},\underline{x})\,
\ket{\Psi^{(e)}_{0}(\underline{R})}
\quad,
\label{eq.vsc_crude_cbo_ground_state}
\end{align}
which we denote as \textit{crude} CBO (CCBO) approximation. The latter equivalently resembles, $\ket{\Psi_0(\underline{R},\underline{x})}\approx \ket{\Psi_\mathrm{ccbo}(\underline{R},\underline{x})}$, where we in particular assume that, $\hat{K}_{\nu\mu}=0$, \textit{i.e.}, all transition dipole NAC elements are neglected. The relevance of the \textit{crude} CBO approximation manifests in an appealing sum-of-products form of the ground state crude cPES, $V_0(\underline{R},\underline{x})$, in Eq.\eqref{eq.crude_vsc_length_cpes} with respect to molecular and cavity degrees of freedom. This representation lies at the heart of effective vibro-polaritonic model Hamiltonians. Thus, according to the arguments presented here, most studies on ground state vibro-polaritonic chemistry do actually not rely on the CBO approximation as formulated in Ref.\cite{flick2017,flick2017cbo}, but its \textit{crude} version as introduced in this work. 

\section{Electron-Photon Entanglement and Crude VSC Perturbation Theory}
\label{sec.entanglement_perturbation}
In previous sections, we have seen that approximate crude CBO and CBO ground state theories differ in certain aspects, which we will discuss now in closer detail. First, based on a reduced density matrix (RDM) analysis, we argue that the crude CBO ground state accounts only for a fraction of electron-photon entanglement relative to the CBO ground state. Second, we introduce a Herzberg-Teller inspired perturbative scheme, denoted as crude VSC perturbation theory (cVSC-PT), which allows us to identify the crude CBO ground state as first-order approximation to the CBO ground state. Higher-order perturbative corrections are subsequently shown to account for non-adiabatic electron-photon correlation effects, which manifest as cavity-induced transition dipole type couplings between adiabatic electronic states.

\subsection{Electron-Photon Entanglement in Approximate CBO Ground States}
\label{subsec.entanglement}
In absence of electron-photon entanglement, an approximate light-matter ground state, $\Psi_0$, is separable in the electron-cavity subspace. Equivalently, the corresponding electronic RDM, $\hat{\rho}^{(e)}$, \textit{does not} contain any vibro-polaritonic contributions.\cite{breuer2007,izmaylov2017} Based on this property, we examine RDM, $\hat{\rho}^{(i)}$, for electrons, nuclei and cavity modes with $i=e,n,c$, as shown in Tab.\ref{tab.rdm_overview} for three approximate ground states: The CBO ground state, $\Psi_\mathrm{cbo}$, the \textit{crude} CBO ground state, $\Psi_\mathrm{ccbo}$, and a partially disentangled reference state, $\Psi_\mathrm{ref}$ (\textit{cf.} Appendix \ref{subsec.rdm_details} for details).
\begin{table}[hbt!]
\caption{Reduced density matrices for electrons, $\hat{\rho}^{(e)}(\underline{r},\underline{r}^\prime)$, nuclei, $\hat{\rho}^{(n)}(\underline{R},\underline{R}^\prime)$, and cavity modes, $\hat{\rho}^{(c)}(\underline{x},\underline{x}^\prime)$, for CBO, $\Psi_\mathrm{cbo}(\underline{r},\underline{R},\underline{x})$, crude CBO, $\Psi_\mathrm{ccbo}(\underline{r},\underline{R},\underline{x})$, and partially disentangled reference states (ref), $\Psi_\mathrm{ref}(\underline{r},\underline{R},\underline{x})$. Vibro-polaritonic overlap integrals, $\tilde{S}^{(i)}_{\mu\nu},\tilde{S}^{(i)}_{00}$, and, $\tilde{s}^{(i)}_{00}$, with $i=n,c$, as well as electronic overlap integrals, $O^{(n)}_{\mu\nu}$, are defined in Appendix \ref{subsec.rdm_details}.}
\renewcommand{\baselinestretch}{1.}
    \centering
    \begin{tabular}{l c c c}
       \hline\hline
          & \quad $\hat{\rho}^{(e)}(\underline{r},\underline{r}^\prime)$ 
          & \quad $\hat{\rho}^{(n)}(\underline{R},\underline{R}^\prime)$ 
          & \quad $\hat{\rho}^{(c)}(\underline{x},\underline{x}^\prime)$  
          \vspace{0.1cm}\\
       \hline
       \vspace{-0.2cm}\\
       $\Psi_\mathrm{cbo}(\underline{r},\underline{R},\underline{x})$ 
       & \quad $\displaystyle\sum_{\mu\nu}\tilde{\rho}^{(e)}_{\mu\nu}$ 
       & \quad $\displaystyle\sum_{\mu\nu}\tilde{S}^{(n)}_{\mu\nu}\,O^{(n)}_{\mu\nu}$ 
       & \quad $\displaystyle\sum_\mu\tilde{S}^{(c)}_{\mu\mu}$  \vspace{0.1cm}\\

       $\Psi_\mathrm{ccbo}(\underline{r},\underline{R},\underline{x})$ 
       & \quad $\tilde{\rho}^{(e)}_{00}$ 
       & \quad $\tilde{S}^{(n)}_{00}\,O^{(n)}_{00}$ 
       & \quad $\tilde{S}^{(c)}_{00}$  \vspace{0.2cm}\\

       $\Psi_\mathrm{ref}(\underline{r},\underline{R},\underline{x})$
       & \quad $\rho^{(e)}_{00}$ 
       & \quad $\tilde{s}^{(n)}_{00}$
       & \quad $\tilde{s}^{(c)}_{00}$ \vspace{0.1cm}\\
       \hline\hline
    \end{tabular}
\label{tab.rdm_overview}
\end{table}

We first recall, that both factors of $\Psi_\mathrm{cbo}=\chi^{(nc)}_0\Psi^{(ec)}_0$ in Eq.\eqref{eq.vsc_cbo_ground_state} depend on nuclear and cavity coordinates. We then realize that the adiabatic electron-photon ground state, $\Psi^{(ec)}_0$, can be expanded in the adiabatic electronic basis, Eq.\eqref{eq.cbo_state_adiabatic_expansion}, which replaces cavity coordinate dependence of $\Psi^{(ec)}_0$ in $\Psi_\mathrm{cbo}$ by contributions from all adiabatic \textit{electronic} states (\textit{cf.} Appendix \ref{subsec.rdm_details}). Thus, CBO-RDM (Tab.\ref{tab.rdm_overview}, first line) are determined by the full adiabatic electronic basis. Furthermore, they contain vibro-polaritonic contributions (indicated by tilde signs), which are traced back to overlap integrals, $\tilde{S}^{(n)}_{\mu\nu}$ and $\tilde{S}^{(c)}_{\mu\mu}$, between nuclear-photon states, $\tilde{\chi}^{(nc)}_\mu$ (\textit{cf.} Appendix \ref{subsec.rdm_details}). 

Crude CBO-RDM (Tab.\ref{tab.rdm_overview}, second line) related to the \textit{crude} CBO ground state, $\Psi_\mathrm{ccbo}=\tilde{\chi}^{(nc)}_0\Psi^{(e)}_0$, in Eq.\eqref{eq.vsc_crude_cbo_ground_state} emerge naturally by truncating adiabatic basis expansions after the first term. This procedure resembles the crude CBO approximation in the wave function picture, which is equivalent to neglecting cavity-induced non-adiabatic transition dipole type couplings (\textit{cf.} Subsec.\ref{subsec.crude_cbo_approx}). We thus argue that electron-photon entanglement in $\Psi_\mathrm{ccbo}$ due to cavity-induced non-adiabatic contributions from excited electronic states is missing relative to $\Psi_\mathrm{cbo}$. 
Further, $\Psi_\mathrm{ccbo}$ is only \textit{apparently} separable in the electron-cavity subspace, since both factors depend on nuclear coordinates, which in turn results in vibro-polaritonic contributions to the electronic crude CBO-RDM, $\tilde{\rho}^{(e)}_{00}$ (\textit{cf.} Appendix \ref{subsec.rdm_details}). From this perspective, electrons and cavity modes are actually not disentangled in $\Psi_\mathrm{ccbo}$, but ``indirectly'' entangled via nuclei. 

Illustratively, the reference state's electronic RDM, $\rho^{(e)}_{00}$, is independent of vibro-polaritonic contributions (Tab.\ref{tab.rdm_overview}, third line), which reflects the fully separable product structure of $\Psi_\mathrm{ref}$, Eq. \eqref{eq.crude_reference_state}, in the electron-cavity subspace: Here, electrons are disentangled with respect to \textit{both} nuclei \textit{and} cavity modes (although the latter are entangled).

\subsection{Ground State VSC Theory from a Perturbative Perspective}
\label{sec.perturbation}
We propose to connect ground state CBO and ground state crude CBO theories via a perturbative scheme, which extends ideas from Herzberg-Teller theory\cite{ballhausen1972} to molecular cQED in the VSC regime. This approach is abbreviated as cVSC-PT($n$) at $n^\mathrm{th}$-order of perturbation theory.

According to Sec.\ref{sec.vsc_theory}, the adiabatic electron-photon ground state, $\ket{\Psi^{(ec)}_0(\underline{R},\underline{x})}$, satisfies the TISE \eqref{eq.vsc_ground_state_tise} with cPES, $\mathcal{E}^{(ec)}_0(\underline{R},\underline{x})=E^{(ec)}_0(\underline{R},\underline{x})$, in the CBO approximation. We now rewrite the electron-photon Hamiltonian as
\begin{align}
\hat{H}_{ec}
&=
\hat{T}_e
+
V(\underline{r},\underline{R},\underline{x}_0)
+
\Delta V(\underline{r},\underline{R},\underline{x})
\quad,
\end{align}
with \textit{difference potential}
\begin{align}
\Delta V(\underline{r},\underline{R},\underline{x})
&=
V(\underline{r},\underline{R},\underline{x})
-
V(\underline{r},\underline{R},\underline{x}_0)
\quad,
\end{align}
and recall that the cavity reference configuration, $\underline{x}_0$, in Eq.\eqref{eq.length_cavity_reference_config} was chosen such that, $\hat{H}_{ec}(\underline{x}_0)=\hat{H}_{e}$. Hence, $V(\underline{r},\underline{R},\underline{x}_0)=V(\underline{r},\underline{R})$, is just the bare molecular potential in Eq.\eqref{eq.molecular_potential} and $\Delta V(\underline{r},\underline{R},\underline{x})$ contains all cavity contributions to $\hat{H}_{ec}$. Subsequently, we apply Rayleigh-Schr\"odinger perturbation theory to approximately solve Eq.\eqref{eq.vsc_ground_state_tise} with zeroth-order electronic Hamiltonian
\begin{align}
\hat{H}_0
&=
\hat{T}_e
+
V(\underline{r},\underline{R},\underline{x}_0)
=
\hat{H}_e
\quad,
\end{align}
and perturbation, $\Delta V(\underline{r},\underline{R},\underline{x})$. Perturbative expansions of the electron-photon ground state and cPES are given by well known expressions
\begin{align}
\ket{\Psi^{(ec)}_0(\underline{R},\underline{x})}
&=
\ket{\Psi^{(0)}_0(\underline{R},\underline{x})}
+
\lambda
\ket{\Psi^{(1)}_0(\underline{R},\underline{x})}
+
\mathcal{O}(\lambda^2)
\;,
\end{align}
and
\begin{multline}
E^{(ec)}_0(\underline{R},\underline{x})
=
E^{(0)}_0(\underline{R},\underline{x})
+
\lambda\,
E^{(1)}_0(\underline{R},\underline{x})
\\
+
\lambda^2\,
E^{(2)}_0(\underline{R},\underline{x})
+
\mathcal{O}(\lambda^3)
\quad.
\end{multline}
At zeroth-order, we find
\begin{align}
\hat{H}_e
\ket{\Psi^{(0)}_0(\underline{R},\underline{x})}
&=
E^{(0)}_0(\underline{R},\underline{x})
\ket{\Psi^{(0)}_0(\underline{R},\underline{x})}
\quad,
\end{align}
which identifies the bare adiabatic electronic ground state, $\ket{\Psi^{(0)}_0(\underline{R},\underline{x})}=\ket{\Psi^{(e)}_0(\underline{R})}$, as zeroth-order state with corresponding \textit{molecular} PES, $E^{(0)}_0(\underline{R},\underline{x})=E^{(e)}_0(\underline{R})$. The first-order corrected cPES is then given by
\begin{align}
V^{(1)}_0(\underline{R},\underline{x})
&=
E^{(e)}_0(\underline{R})
+
\underbrace{\braket{
\Psi^{(e)}_0
\vert
\Delta V
\vert
\Psi^{(e)}_0}_{\underline{r}}}_{=E^{(1)}_0(\underline{R},\underline{x})}
\quad,
\end{align}
where the difference potential, $\Delta V$, reads
\begin{align}
\Delta V
&=
V_c
+
\hat{H}_{sc}
+
\hat{H}_{dse}
\quad,
\label{eq.difference_pot_length}
\end{align}
such that we recover for, $V^{(1)}_0(\underline{R},\underline{x})$, the corresponding ground state \textit{crude} cPES, Eq.\eqref{eq.crude_vsc_length_cpes}. Since $E^{(1)}_0(\underline{R},\underline{x})$ is fully determined by the zeroth-order wave function, \textit{i.e.}, the adiabatic electronic ground state, in line with the crude CBO ground state in Eq.\eqref{eq.vsc_crude_cbo_ground_state}, ground state \textit{crude} VSC theory can be interpreted as first-order approximation to ground state VSC theory, denoted by cVSC-PT(1). 

Non-adiabatic corrections accounting for electron-photon correlation enter in cVSC-PT(2) as
\begin{align}
V^{(2)}_0(\underline{R},\underline{x})
&=
V^{(1)}_0(\underline{R},\underline{x})
+
\underbrace{
\sum_{\mu\neq0}
\dfrac{
\vert
\Delta V_{0\mu}(\underline{R},\underline{x})
\vert^2}
{\Delta E^{(e)}_{0\mu}(\underline{R})}}_{=E^{(2)}_0(\underline{R},\underline{x})}
\quad,
\end{align}
with, $\Delta V_{0\mu}=\braket{\Psi^{(e)}_0\vert\Delta V\vert\Psi^{(e)}_\mu}_{\underline{r}}$, and, $\Delta E^{(e)}_{0\mu}=E^{(e)}_0-E^{(e)}_\mu$, as determined by the first-order corrected state
\begin{align}
\ket{\Psi^{(1)}_0(\underline{R},\underline{x})}
&=
\sum_{\mu\neq0}
\dfrac{
\Delta V_{0\mu}(\underline{R},\underline{x})}
{\Delta E^{(e)}_{0\mu}(\underline{R})}\,
\ket{\Psi^{(e)}_\mu(\underline{R})}
\quad.
\end{align}
In order to illustrate how electron-photon correlation corrections manifest perturbatively, we discuss $E^{(2)}_0(\underline{R},\underline{x})$ in more detail. From Eq.\eqref{eq.difference_pot_length}, we recognize the second-order correction to be independent of the bare cavity potential, since matrix elements, $\Delta V_{0\mu}$, are strictly off-diagonal. Hence, we explicitly have
\begin{multline}
E^{(2)}_0(\underline{R},\underline{x})
=
\sum_{\mu\neq0}
\dfrac{
\vert
H^{sc}_{0\mu}
\vert^2
+
2\,H^{sc}_{0\mu}\,H^{dse}_{0\mu}
+
\vert
H^{dse}_{0\mu}
\vert^2
}
{\Delta E^{(e)}_{0\mu}(\underline{R})}
\,,
\label{eq.length_2nd_order_energy}
\end{multline}
with matrix element, $H^{sc}_{0\mu}=\braket{\Psi^{(e)}_0\vert\hat{H}_{sc}\vert\Psi^{(e)}_\mu}_{\underline{r}}$ (equivalently for, $H^{dse}_{0\mu}$). Further, we recall scaling with respect to the light-matter interaction constant as, $\hat{H}_{sc}\propto g$, and, $\hat{H}_{dse}\propto g^2$ (we dropped mode indices for brevity), such that, $\vert H^{sc}_{0\mu}\vert^2\propto g^2$, in Eq.\eqref{eq.length_2nd_order_energy} provides the leading-order contribution to $E^{(2)}_0(\underline{R},\underline{x})$, which reads 
\begin{multline}
\vert H^{sc}_{0\mu}\vert^2
=
\sum^{2N_c}_{\lambda,k}
\sum^{2N_c}_{\lambda^\prime,k^\prime}
\sqrt{\dfrac{2\omega_k}{\hbar}}
\sqrt{\dfrac{2\omega_{k^\prime}}{\hbar}}\,
g_k\,
g_{k^\prime}
\\
\times
\biggl(
\underline{e}_{\lambda k}
\cdot
\underline{d}_{0\mu}
\biggr)
\biggl(
\underline{e}_{\lambda^\prime k^\prime}
\cdot
\underline{d}_{0\mu}
\biggr)\,
x_{\lambda k}\,
x_{\lambda^\prime k^\prime}
\quad.
\end{multline}
This expression accounts for correlations between electrons and cavity modes via a cavity-induced transition-dipole type interaction between adiabatic electronic states, which in turn couples cavity modes with potentially distinct mode and polarization indices, $k,k^\prime$, and, $\lambda,\lambda^\prime$. Interestingly, a similar observation has been recently reported from a non-perturbative perspective.\cite{bonini2022}

\section{Numerical Analysis}
\label{sec.numerical_analysis}
We now numerically analyse several aspects of VSC and crude VSC theory for the cavity Shin-Metiu (CSM) model\cite{flick2017}. The CSM model provides a numerically exactly solvable model, which has been proven versatile to capture the rich physics of non-adiabatic phenomena involving moving electrons and nuclei coupled to quantized cavity modes.

\subsection{The Cavity Shin-Metiu Model}
\begin{figure}[hbt]
\begin{center}
\includegraphics[scale=1.0]{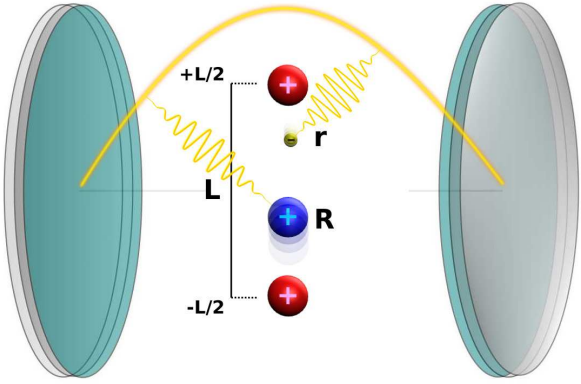}
\end{center}
\renewcommand{\baselinestretch}{1.}
\caption{Schematic representation of the cavity Shin Metiu model with a moving positively charged nucleus (blue) with coordinate, $R$, and a moving negatively charged electron (yellow) with coordinate, $r$, both interacting with a single cavity mode. Fixed positively charged nuclei are colored in red and located at a distance $L$ from each other, while mobile particles move along the related molecular axis connecting fixed nuclei.} 
\label{fig.cavity_shin_metiu}
\end{figure}

The cavity Shin-Metiu (CSM) model describes a ``molecular'' model system composed of a moving nucleus and a moving electron\cite{shinmetiu1995}, which couple via their charges to a single quantized cavity mode\cite{flick2017} as schematically depicted in Fig.\eqref{fig.cavity_shin_metiu}. Both mobile particles move along a molecular axis connecting two fixed nuclei at a given distance, $L$. 

The length-gauge Pauli-Fierz Hamiltonian, Eq.\eqref{eq.length_pauli_fierz_short}, reads for the CSM model (in atomic units)
\begin{align}
\begin{matrix}
\hat{T}_\mathrm{n}
+
\hat{T}_\mathrm{c}
=
-
\dfrac{1}{2M_a}
\dfrac{\partial^2}{\partial R^2}
-
\dfrac{1}{2}
\dfrac{\partial^2}{\partial x^2_c}
\vspace{0.2cm}
\\
\hat{H}_{ec}
=
\hat{H}_e
+
\dfrac{\omega^2_c}{2}
\left(
x_c
+
\sqrt{\dfrac{2}{\omega^3_c}}\,
g_c\,
d_{en}(r,R)
\right)^2
\vspace{0.2cm}
\\
\hat{H}_e
=
-
\dfrac{1}{2}
\dfrac{\partial^2}{\partial r^2}
+
V(r,R)
\end{matrix}
\quad,
\label{eq.length_csm_hamiltonian}
\end{align}
with single-mode cavity displacement coordinate, $x_c$, molecular dipole moment
\begin{align}
d_{en}(r,R)
&=
-
r
+
Q_a R
\quad,
\end{align}
and Shin-Metiu potential\cite{shinmetiu1995}
\begin{align}
V(r,R)
=
V_n(R)
+
V_{en}(r,R)
\quad.
\end{align}
Here, the first term relates to a purely repulsive Coulomb interaction between the moving and fixed nuclei positioned at $\pm L/2$ 
\begin{align}
V_n(R)
&=
\dfrac{Q_a\,Q}{\vert L/2-R\vert}
+
\dfrac{Q_a\,Q}{\vert L/2+R\vert}
\quad,
\end{align}
with nuclear coordinate, $R$, and nuclear charges, $Q_a=Q=1$, respectively. The attractive ``softened'' Coulomb interaction between the electron and all nuclei reads
\begin{multline}
V_{en}(r,R)
=
-
\dfrac{Q\,\mathrm{erf}\left(\dfrac{\vert L/2 - r\vert}{R_f}\right)}{\vert L/2 - r\vert}
\\
-
\dfrac{Q\,\mathrm{erf}\left(\dfrac{\vert L/2 + r\vert}{R_f}\right)}{\vert L/2 + r\vert}
-
\dfrac{Q_a\,\mathrm{erf}\left(\dfrac{\vert R - r\vert}{R_c}\right)}{\vert R - r \vert}
\quad,
\end{multline}
with error function, $\mathrm{erf}(\dots)$, besides screening lengths for fixed, $R_f$, and moving nuclei, $R_c$. The latter determines the nuclear non-adiabatic coupling strength.\cite{shinmetiu1995} Following earlier work\cite{flick2017cbo}, we consider a model with $L=10\,\text{\AA}$ and $R_f=1.5\,\text{\AA}$, where we set, $R_c=1.5\,\text{\AA}$, for a ``weak'' and, $R_c=2.0\,\text{\AA}$, for a ``strong'' non-adiabatic coupling. 

CSM model parameters, which determine the VSC scenario, are obtained by first solving the electronic TISE with $\hat{H}_e$ given in Eq.\eqref{eq.length_csm_hamiltonian}
\begin{align}
\hat{H}_e\,
\Psi^{(e)}_0(r;R)
&=
E^{(e)}_0(R)\,
\Psi^{(e)}_0(r;R)
\quad,
\label{eq.electronic_sm}
\end{align}
for the ground state molecular PES, $E^{(e)}_0(R)$, and subsequently the corresponding nuclear TISE 
\begin{align}
\left(
\hat{T}_n
+
E^{(e)}_0(R)
\right)
\varphi^{(n)}_v(R)
&=
\varepsilon_v\,
\varphi^{(n)}_v(R)
\quad,
\label{eq.nuclear_sm}
\end{align}
for vibrational eigenstates, $\varphi^{(n)}_v(R)$, with energies, $\varepsilon_v$. For the CSM model under VSC, we tune the cavity resonant to the fundamental vibrational transition, $\hbar\omega_c=\varepsilon_3-\varepsilon_0$. For the weak (strong) NAC regime, we have a cavity frequency, $\hbar\omega_c=585.16\,\mathrm{cm}^{-1}\,(516.74\,\mathrm{cm}^{-1})$, with vibrational transition dipole element, $d_{30}=0.089\,e a_0\,(0.26\,e a_0)$. The light-matter interaction constant, $g=\frac{\hbar\omega_c}{d_{30}}\,\eta$, is furthermore characterized by a dimensionless parameter, $\eta$.\cite{kockum2019} Further model parameters and numerical details on the CSM model are provided in Appendix \ref{subsec.numerics_csm_model}. 

\subsection{Adiabatic Electron-Photon States}
\begin{figure*}[hbt]
\begin{center}
\includegraphics[scale=1.0]{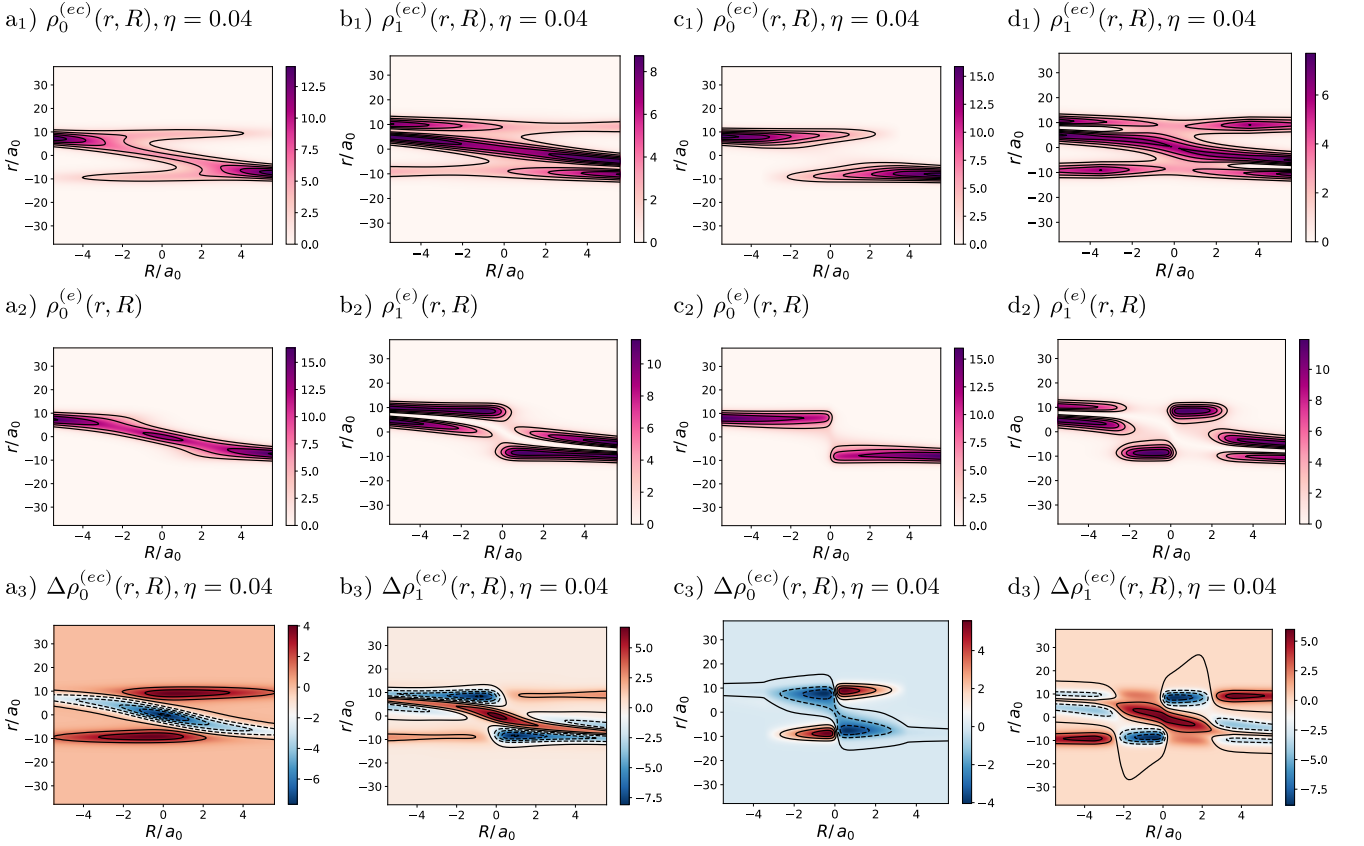}
\end{center}
\renewcommand{\baselinestretch}{1.}
\caption{Contour plots of reduced adiabatic electron-photon densities, $\rho^{(ec)}_\nu(r,R)$, (a$_1$-d$_1$), reduced adiabatic electron densities, $\rho^{(e)}_\nu(r,R)$, (a$_2$-d$_2$) and adiabatic difference densities, $\Delta \rho^{(ec)}_\nu(r,R)$, (a$_3$-d$_3$) in electron-nuclear coordinate space for adiabatic ground and first excited states ($\nu=0,1$) in weak and strong non-adiabatic coupling (NAC) regime of the CSM model under vibrational strong coupling with $\eta=0.04$.} 
\label{fig.adiabatic_densities_length}
\end{figure*}
We begin our discussion by examining the impact of vibrational strong coupling on adiabatic electron-photon states, $\Psi^{(ec)}_\nu$. We restrict ourselves to the respective ground and first excited states of the CSM model with quantum numbers, $\nu=0,1$, and consider reduced densities in the molecular coordinate subspace
\begin{align}
\rho^{(ec)}_\nu(r,R)
&=
\displaystyle\int
\mathrm{d}x_c\,
\vert
\Psi^{(ec)}_\nu(r;R,x_c)
\vert^2
\quad,
\vspace{0.2cm}
\\
\rho^{(e)}_\nu(r,R)
&=
\vert
\Psi^{(e)}_\nu(r;R)
\vert^2
\quad,
\end{align}
where, $\Psi^{(ec)}_\nu(r;R,x_c)$, are eigenstates of $\hat{H}_{ec}$ in Eq.\eqref{eq.length_csm_hamiltonian}. In order to access cavity-induced changes in reduced light-matter hybrid densities, $\rho^{(ec)}_\nu$, relative to the bare molecular densities, $\rho^{(e)}_\nu$ (determined by adiabatic electronic states, $\Psi^{(e)}_\nu$), we introduce a state-dependent difference density\cite{flick2017cbo} 
\begin{align}
\Delta \rho^{(ec)}_\nu(r,R)
&=
\rho^{(ec)}_\nu(r,R)
-
\rho^{(e)}_\nu(r,R)
\quad,
\end{align}
which vanishes in the non-interacting limit, \textit{i.e.}, $\Delta \rho^{(ec)}_\nu(r,R)=0$, for, $\eta=0$. 

In Fig.\ref{fig.adiabatic_densities_length}, $\rho^{(ec)}_\nu(r,R)$, $\rho^{(e)}_\nu(r,R)$ and $\Delta \rho^{(ec)}_\nu(r,R)$ are shown for VSC with $\eta=0.04$, and quantum numbers $\nu=0,1$ for both the weak and strong non-adiabatic coupling regimes. We observe cavity-induced density redistribution in both the weakly and the strongly non-adiabatic CSM model for the ground and first excited electron-photon state: Red coloured regions in $\Delta \rho^{(ec)}_\nu(r,R)$ indicate a density increase and blue coloured regions a density decrease. In the weakly non-adiabatic regime, new maxima emerge along the electron coordinate, $r$, while density is dominantly shifted along the nuclear coordinate, $R$, under strong non-adiabatic coupling. 

Density variations with $\eta$ indicate non-trivial electron-photon correlation effects in energetically low lying adiabatic electron-photon states of the CSM model, despite the fact that cavity photon energies lie significantly below characteristic electronic excitation energies. Based on those observations, an approximation of the adiabatic electron-photon ground state by the bare electronic ground state, \textit{i.e.}, $\rho^{(ec)}_0(r,R)\approx\rho^{(e)}_0(r,R)$ (crude CBO approximation), does therefore not necessarily seem to be an always well justified simplification. We will provide further arguments supporting this statement in Subsec.\ref{subsec.electron_photon_corr}.

\subsection{Nuclear and Cavity Non-Adiabatic Coupling}
We now turn to non-adiabatic coupling under VSC at $\eta=0.04$ and discuss both nuclear and cavity derivative NAC elements for weak and strong NAC regimes of the CSM model. 
\begin{figure*}[hbt!]
\begin{center}
\includegraphics[scale=1.0]{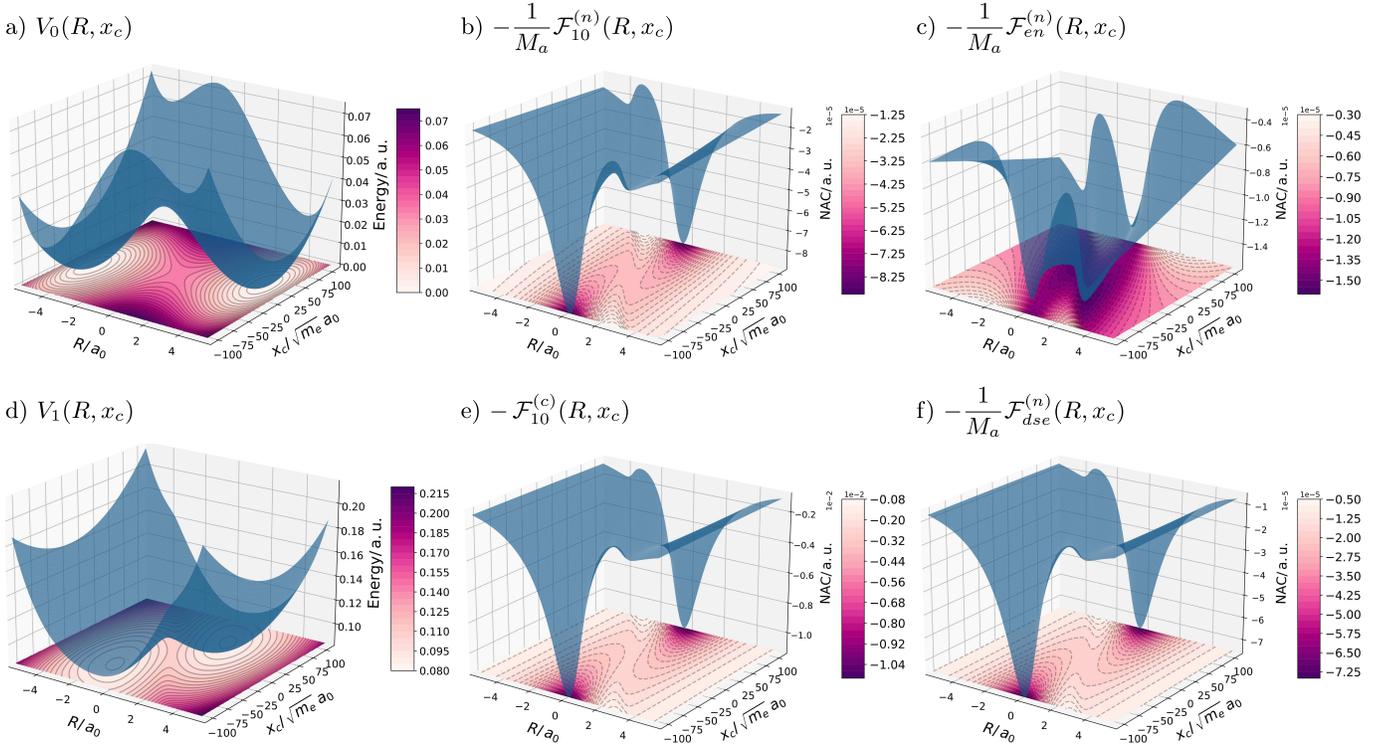}
\end{center}
\renewcommand{\baselinestretch}{1.}
\caption{Weakly non-adiabatic CSM model under VSC with $\eta=0.04$: a) Ground, $V_0(R,x_c)$, and d) first excited state cPES, $V_1(R,x_c)$, as function of nuclear, $R$, and cavity displacement, $x_c$, coordinates with energies in atomic units (a.u.). b) Mass-weighted nuclear derivative NAC element, $-\frac{1}{M_a}\mathcal{F}^{(n)}_{10}(R,x_c)$, and e) cavity derivative NAC element, $-\,\mathcal{F}^{(c)}_{10}(R,x_c)$, in a.u. as function of coordinates under VSC. c) Molecular, $-\frac{1}{M_a}\mathcal{F}^{(n)}_{en}(R,x_c)$, and f) DSE-induced nuclear NAC contributions, $-\frac{1}{M_a}\mathcal{F}^{(n)}_{dse}(R,x_c)$, to mass-weighted nuclear NAC element in b) with same parameters.} 
\label{fig.vsc_nac_length}
\end{figure*}

As observed before in the weak non-adiabatic coupling regime\cite{flick2017cbo}, both the ground and first excited state cPES, $V_0(R,x_c)$ and $V_1(R,x_c)$, are distorted in the $(R,x_c)$-plane under VSC (\textit{cf.} Figs.\ref{fig.vsc_nac_length}a and \ref{fig.vsc_nac_length}d). Both states are subject to nuclear (mass-weighted) and cavity derivative NAC elements, $\mathcal{F}^{(n)}_{10}$ and $\mathcal{F}^{(c)}_{10}$, which are found to behave qualitatively similar. According to Figs.\ref{fig.vsc_nac_length}b and \ref{fig.vsc_nac_length}e, two dominant minima with strongest NACs are observed at positions, where the distorted cPES are energetically close. Minima are located close to the nuclear coordinate's origin and take values of $x_c\approx \pm100\,\sqrt{m_e}\,a_0$ for the cavity displacement coordinate corresponding to roughly, $n_c=13$, photons as estimated from classical turning points of the harmonic cavity potential. The cavity NAC element for the CSM model
\begin{align}
\mathcal{F}^{(c)}_{10}(R,x_c)
&=
\dfrac{\sqrt{2\,\hbar\omega^3_c}}{d_{30}}\,
\eta\,
\dfrac{\mathcal{D}_{10}(R,x_c)}{\Delta E^{(ec)}_{10}(R,x_c)}
\quad,
\end{align}
is observed to be two orders of magnitude larger compared to the nuclear derivative NAC, $\mathcal{F}^{(n)}_{10}$. Here, $d_{30}$, is the vibrational transition dipole moment, whereas $\Delta E^{(ec)}_{10}$ relates to the energy difference between adiabatic electron-photon states. 

The nuclear term, $\mathcal{F}^{(n)}_{10}$, has according to Eq.\eqref{eq.length_nuclear_nac_hellmann_feynman} a vibronic, $\mathcal{F}^{(n)}_{en}$, and a cavity-induced contribution, $\mathcal{F}^{(n)}_{dse}$, which here explicitly read
\begin{align}
\mathcal{F}^{(n)}_{en}(R,x_c)
&=
\dfrac{\braket{\Psi^{(ec)}_1(R,x_c)
\vert
\dfrac{\partial\,V_{en}}{\partial R}
\vert
\Psi^{(ec)}_0(R,x_c)}_{r}}{\Delta E^{(ec)}_{10}(R,x_c)}
\,,
\label{eq.nuc_nac_length_molecule}
\vspace{0.2cm}
\\
\mathcal{F}^{(n)}_{dse}(R,x_c)
&=
\frac{2\,Q_a\,\eta^2}{d_{30}}\,\dfrac{\mathcal{D}^{(ec)}_{10}(R,x_c)}{\Delta E^{(ec)}_{10}(R,x_c)}
\quad.
\label{eq.nuc_nac_length_dse}
\end{align}
From Figs.\ref{fig.vsc_nac_length}c and \ref{fig.vsc_nac_length}f, we observe the DSE-induced term, $\mathcal{F}^{(n)}_{dse}$, to be roughly one order of magnitude larger compared to the vibronic contribution, $\mathcal{F}^{(n)}_{en}$, under VSC with $\eta=0.04$. In particular, the DSE contribution, $\mathcal{F}^{(n)}_{dse}$, is qualitatively similar to $\mathcal{F}^{(c)}_{10}$, which can be traced back to the \textit{bright} transition dipole moment, $\mathcal{D}^{(ec)}_{10}(R,x_c)$. The difference in magnitude is explained through different scaling with respect to $\eta$. 
Consequently, nuclear derivative NAC elements turn out to be significantly influenced by light-matter interaction, when transitions between two adiabatic states have a strong \textit{bright} component (\textit{cf.} Subsec.\ref{subsec.length_nacs_details}). This finding contrasts the bare molecular picture, where optical properties of adiabatic states are not relevant for non-adiabatic coupling. Moreover, since cavity NAC contributions are here found to be larger in magnitude than their molecular counterparts, the notion of a \textit{weak} NAC regime does interestingly not simply transfer from the molecular to the light-matter hybrid scenario, at least in the herein presented model.
\begin{figure*}[hbt]
\begin{center}
\includegraphics[scale=1.0]{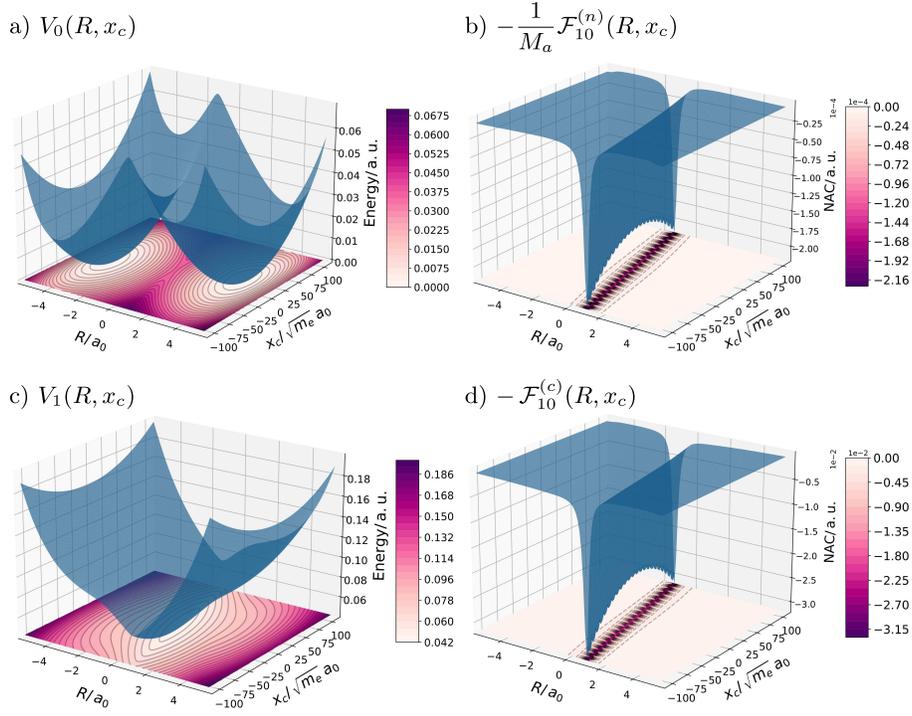}
\end{center}
\renewcommand{\baselinestretch}{1.}
\caption{Strongly non-adiabatic CSM model under VSC with $\eta=0.04$: a) Ground, $V_0(R,x_c)$, and c) first excited state cPES, $V_1(R,x_c)$, as function of nuclear, $R$, and cavity displacement, $x_c$, coordinates with energies in atomic units (a.u.). b) Mass-weighted nuclear derivative NAC element, $-\frac{1}{M_a}\mathcal{F}^{(n)}_{10}(R,x_c)$, and d) cavity derivative NAC element, $-\,\mathcal{F}^{(c)}_{10}(R,x_c)$, in a.u. as function of coordinates under VSC as a) and d).} 
\label{fig.vsc_strong_nac_length}
\end{figure*}

For the strong NAC regime under VSC, cPESs and derivative NAC elements are presented in Fig.\ref{fig.vsc_strong_nac_length}. The ground state cPES shows here a pronounced double minimum, whereas the energetically close lying first excited state exhibits a single minimum close to the ground state's transition state.\cite{flick2017cbo} NAC elements exhibit a sharp maximum with evenly spaced, large amplitude peaks along a seam in the $(R,x_c)$-plane, which is parallel to the separating surface between the two minima of the ground state cPES. A similar observation has been reported very recently.\cite{schnappinger2023} The maximal amplitude is naturally increased compared to the weak NAC regime and the cavity contribution is again roughly two orders of magnitude larger than the vibronic coupling. This aspect points at another difference in the notion of \textit{weak} and \textit{strong} NAC regimes compared between light-matter hybrid systems and the bare molecular picture: For the system studied here, \textit{cavity} weak non-adiabatic coupling is characterized by relatively localized peaks in NAC elements, \textit{i.e.}, where cPESs come close, which only becomes relevant when the system is excited in the cavity subspace. In contrast, \textit{cavity} strong non-adiabatic coupling exhibits substantially extended non-adiabatic interactions between ground and first excited state cPESs as soon as the hybrid system enters the transition state region, irrespective of the energy stored in the cavity mode.

\subsection{Electron-Photon Correlation and Cavity PES}
\label{subsec.electron_photon_corr}
Cavity potential energy surfaces are conceptually appealing, since they directly generalize well established ideas from quantum chemistry to the realm of vibro-polaritonic chemistry. Hence, common quantum chemical concepts of (classical) activation barriers\cite{li2021a,fischer2022a,sun2022,lindoy2022a} and minimum energy paths (MEP)\cite{miller1980,kraka2011,fischer2022a} are naturally transferred to mechanistic arguments in thermal vibro-polaritonic chemistry. However, correlated CBO and crude CBO approaches are commonly not distinguished, which motivates us to discuss how the different treatment of electron-photon correlation impacts those properties. 
\begin{figure}[hbt!]
\begin{center}
\includegraphics[scale=1.0]{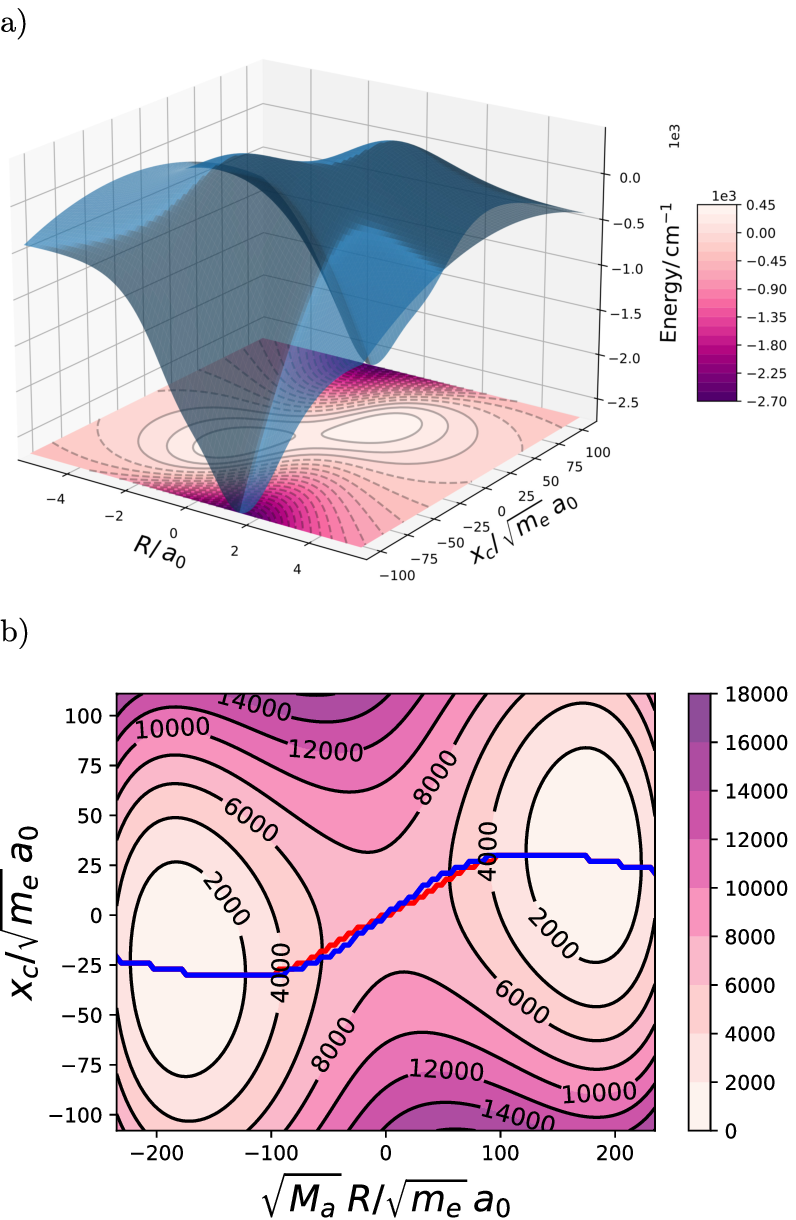}
\end{center}
\renewcommand{\baselinestretch}{1.}
\caption{Differences between ground state cPES and ground state crude cPES for the weakly non-adiabatic CSM model under VSC with $\eta=0.04$. (a) Energy difference, $\Delta E_0(R,x_c)$, in wave numbers $(\mathrm{cm}^{-1})$ and (b) numerical approximations to cavity minimum energy paths for $E^{(ec)}_0(R,x_c)$ in red and for $V_0(R,x_c)$ in blue (contour plot shown in $\mathrm{cm}^{-1}$). } 
\label{fig.vsc_minimum_energy_path_length}
\end{figure}

In Fig.\ref{fig.vsc_minimum_energy_path_length}a, we show the energy difference between the ground state cPES and its crude counterpart introduced in Eq.\eqref{eq.crude_vsc_length_cpes} under VSC at $\eta=0.04$ 
\begin{align}
\Delta E_0(R,x_c)
&=
E^{(ec)}_0(R,x_c)
-
V_0(R,x_c)
\end{align}
for the weakly non-adiabatic CSM model. We find pronounced maxima in $\Delta E_0$ close to the transition state along the nuclear coordinate and negative regions along the cavity displacement coordinate. The transition state is determined by classical activation barriers, which are defined as the energy difference between the global minimum and the transition state energy
\begin{align}
\begin{matrix}
\Delta \mathcal{E}^a_\mathrm{cl}
=
E^{(ec)}_0(R^\ddagger,x^\ddagger_c)
-
\mathrm{min}\,E^{(ec)}_0(R,x_c)
\vspace{0.2cm}
\\
\Delta E^a_\mathrm{cl}
=
V_0(R^\ddagger,x^\ddagger_c)
-
\mathrm{min}\,V_0(R,x_c)
\end{matrix}
\quad,
\end{align}
for both CBO and crude CBO ground state cPES. In Tab.\ref{tab.activation_energies}, classical activation barriers are shown for the ground state cPES of the CSM model for different light-matter interaction strengths, $\eta$, as obtained under the crude CBO approximation, $\Delta E^a_\mathrm{cl}$, and the CBO approximation, $\Delta\mathcal{E}^a_\mathrm{cl}$. We observe classical activation barriers in crude VSC theory, $\Delta E^a_\mathrm{cl}$, to be nearly independent of $\eta$ for the values studied here, which is in line with results reported in literature, fully accounting for the DSE term.\cite{fischer2021,li2021a} In contrast, classical barriers obtained from fully correlated calculations, $\Delta\mathcal{E}^a_\mathrm{cl}$, increase with $\eta$, here by about $4\,\%$. We note, this can be already qualitatively observed in Ref.\cite{flick2017cbo}. 
\begin{table}[hbt!]
    \centering
    \begin{tabular}{l | r r r r r }
       \hline\hline
         $\eta$ & \quad $0.0$ & \quad $0.01$ & \quad   $0.02$ & \quad   $0.03$  & \quad  $0.04$ \vspace{0.1cm}\\
       \hline
       \vspace{0.1cm}
       $\Delta E^a_\mathrm{cl}/\,\mathrm{cm}^{-1}$ & $7054$ & $7054$ & $7053$ & $7053$ & $7053$ \vspace{0.1cm}\\
       $\Delta\mathcal{E}^a_\mathrm{cl}/\,\mathrm{cm}^{-1}$ & $7054$ &  $7080$ & $7152$ & $7246$ & $7340$ \vspace{0.2cm}\\
       $\Delta E^{(2)}_\mathrm{cl}/\,\mathrm{cm}^{-1}$ & $7054$ & $7053$  & $7047$ & $7031$ & $6994$ \vspace{0.1cm}\\ 
       $\Delta E^{(3)}_\mathrm{cl}/\,\mathrm{cm}^{-1}$ & $7054$ & $7053$  & $7055$ & $7063$ & $7164$ \vspace{0.1cm}\\   
       $\Delta E^{(2,1)}_\mathrm{cl}+\mathcal{O}(\eta^3)/\,\mathrm{cm}^{-1}$ & $7054$ & $7053$  & $7057$ & $7071$ & $7116$ \vspace{0.1cm}\\     
       \hline\hline
    \end{tabular}
\renewcommand{\baselinestretch}{1.}
\caption{Classical activation barriers on the ground state cPES of the weakly non-adiabatic CSM model as obtained from crude VSC theory, $\Delta E^a_\mathrm{cl}$, VSC theory, $\Delta\mathcal{E}^a_\mathrm{cl}$, cVSC-PT(2), $\Delta E^{(2)}_\mathrm{cl}$, cVSC-PT(3), $\Delta E^{(3)}_\mathrm{cl}$, and leading order cVSC-PT(2), $\Delta E^{(2,1)}_\mathrm{cl}$, including only quadratic corrections in $\eta$ (\textit{cf.} Subsec.\ref{subsec.perturb_corr_cpes}). All energies are given in wave numbers for selected values of $\eta$. 
}
\label{tab.activation_energies}
\end{table}

Hence, electron-photon correlation can lead to cavity-induced changes in classical activation barriers on CBO cPESs, which in turn will potentially influence the reactive system's kinetics. This observation is naturally assumed to be system dependent. Further, since crude cPESs do not exhibit a light-matter interaction dependent classical activation barrier, a comparison to CBO cPESs should be taken with care. However, crude cPESs studied here are indeed able to capture some light-matter interaction effects, \textit{e.g.}, barrier broadening and transition state valley narrowing.\cite{li2021a,fischer2022a} 

Eventually, we consider differences in the cavity minimum energy path (MEP), another concept related to reaction kinetics beyond transition state arguments only: Along the (cavity) MEP, the energy is minimized with respect to variations along all orthogonal \textit{valley} coordinates.\cite{miller1980} In Fig.\ref{fig.vsc_minimum_energy_path_length}b, we compare cavity MEPs obtained for both ground state crude and CBO cPES at $\eta=0.04$ and observe a rather close match, \textit{i.e.}, the path's curvature is qualitatively similar on CBO and crude cPES: Both MEPs show a linear behaviour close to the transition state, whereas the crude version is subject to a slightly stronger curvature. Accordingly, electron-photon correlation seems to be less important for the cavity MEP.

\subsection{Perturbatively Corrected Ground State Crude cPES}
\label{subsec.perturb_corr_cpes}
In order to improve the crude description without having to apply exact VSC theory, we now examine perturbative corrections to the ground state crude cPES based on the cVSC-PT formalism introduced in Subsec.\ref{sec.perturbation}. We discuss second- and third-order corrected cPESs of the weakly non-adiabatic CSM model in the CBO approximation, as given by
\begin{align}
V^{(2)}_0(\underline{R},\underline{x})
&=
V^{(1)}_0(\underline{R},\underline{x})
+
E^{(2)}_0(R,x_c)
\quad,
\end{align} 
and
\begin{align}
V^{(3)}_0(\underline{R},\underline{x})
&=
V^{(2)}_0(\underline{R},\underline{x})
+
E^{(3)}_0(R,x_c)
\quad,
\end{align} 
where, $E^{(3)}_0$, is explicitly given in Appendix \ref{subsec.third_order_cpes_correction}. According to Eq.\eqref{eq.length_2nd_order_energy}, the second-order correction decomposes into three contributions, $E^{(2)}_0=\sum^3_{i=1}E^{(2)}_{0,i}$, which read for the CSM model 
\begin{align}
E^{(2)}_{0,1}
&=
\sum_{\mu\neq0}
\dfrac{\vert H^{sc}_{0\mu}\vert^2}
{\Delta E^{(e)}_{0\mu}}
=
\sum_{\mu\neq0}
\dfrac{
2\omega_c\,
g^2
d^2_{0\mu}
x^2_c}
{\Delta E^{(e)}_{0\mu}}
\quad,
\label{eq.cvscpt_2_int_int}
\vspace{0.2cm}
\\
E^{(2)}_{0,2}
&=
\sum_{\mu\neq0}
\dfrac{2\,H^{sc}_{0\mu}\,H^{dse}_{\mu 0}}
{\Delta E^{(e)}_{0\mu}}
\quad,
\nonumber
\vspace{0.2cm}
\\
&=
\sum_{\mu\neq0}
\sum_\alpha
\sqrt{\dfrac{8}{\omega_c}}
\dfrac{g^3
d_{0\mu}
\left(
d_{\mu\alpha}
d_{\alpha0}
\right)
x_c}
{\Delta E^{(e)}_{0\mu}}
\quad,
\vspace{0.2cm}
\\
E^{(2)}_{0,3}
&=
\sum_{\mu\neq0}
\dfrac{\vert H^{dse}_{0\mu}\vert^2}
{\Delta E^{(e)}_{0\mu}}
\quad,
\nonumber
\vspace{0.2cm}
\\
&=
\sum_{\mu\neq0}
\sum_{\alpha,\beta}
\dfrac{g^4}{\omega^2_c}\,
\dfrac{
\left(d_{0\alpha}d_{\alpha\mu}\right)
\left(d_{\mu\beta}d_{\beta 0}\right)
}
{\Delta E^{(e)}_{0\mu}}
\quad,
\end{align}
where, $g=\frac{\hbar\omega_c}{d_{30}}\,\eta$. We suppress nuclear and cavity coordinate dependence for brevity and indicate contributions from DSE expectation values by brackets, \textit{e.g.}, $\left(d_{0\alpha}d_{\alpha\mu}\right)$. Numerically converged results are obtained via inclusion of all neutral states of the weakly non-adiabatic CSM model at fixed $\eta$. 

We illustratively discuss the sensitivity of classical activation barriers to electron-photon correlation effects due to their potential relevance in thermal vibro-polaritonic chemistry. In Tab.\ref{tab.activation_energies}, second-order, $\Delta E^{(2)}_\mathrm{cl}$, and third-order, $\Delta E^{(3)}_\mathrm{cl}$, corrected classical activation barriers of the CSM model are provided for selected values of $\eta$. We immediately recognize sizeable differences with respect to the crude barrier energy, $\Delta E^{a}_\mathrm{cl}$, which points at the relevance of beyond-crude CBO electron-photon correlation in low-frequency cavity mode settings. While the second-order correction leads to a reduction of the barrier, we recover the correct trend and a significant fraction of barrier energy increase relative to the fully correlated result at third-order.

Finally, we note a significantly corrected classical activation barrier, $\Delta E^{(2,1)}_\mathrm{cl}$, when accounting only for the leading order term of cVSC-PT(2) in Eq.\eqref{eq.cvscpt_2_int_int}. Such a result is very appealing since the numerical effort towards a good approximation is rather low. However, this correction has to be treated carefully, since it always leads to a barrier \textit{increase}. This statement is rationalized by noting that $E^{(2)}_{0,1}$ vanishes at the transition state located at $x_c=0.0$, since only there the cavity coordinate gradient tends to zero as required by the minimum energy path. Further, since the numerator is strictly positive, the denominator leads to a negative $E^{(2)}_{0,1}$, which indicates an energy decrease at the reactant configuration and therefore an effective barrier \textit{increase}. In this context, it is an interesting open question, which we here leave for future studies, whether classical activation barriers on cPES can actually be \textit{lowered} due to light-matter interaction and under which circumstance this scenario appears to happen. 
\subsection{The Electron-Polariton Hamiltonian in Vibro-Polaritonic Chemistry}
\label{subsec.elec_pol_gs}
In this last section, we discuss potential issues occurring when the \textit{electron-polariton} Hamiltonian, $\hat{H}_p=\hat{T}_c+\hat{H}_{ec}$, is employed in the VSC regime (\textit{cf.} Appendix \ref{subsec.polariton_hamiltonian} for details). In $\hat{H}_p$, the electron-photon Hamiltonian, $\hat{H}_{ec}$, is augmented by the cavity KEO, $\hat{T}_c$, which motivates its applicability in the electronic strong coupling regime\cite{haugland2020}: Here, both electrons and cavity modes are treated as ``fast'' degrees of freedom due to similar energy scales opposed to ``slow'' nuclei. Motivated by a recent study\cite{riso2022}, we will now analyse a VSC scenario for the CSM model, where we employ, $\hat{H}_p$, to calculate cPESs.

\begin{figure*}[hbt]
\begin{center}
\includegraphics[scale=1.0]{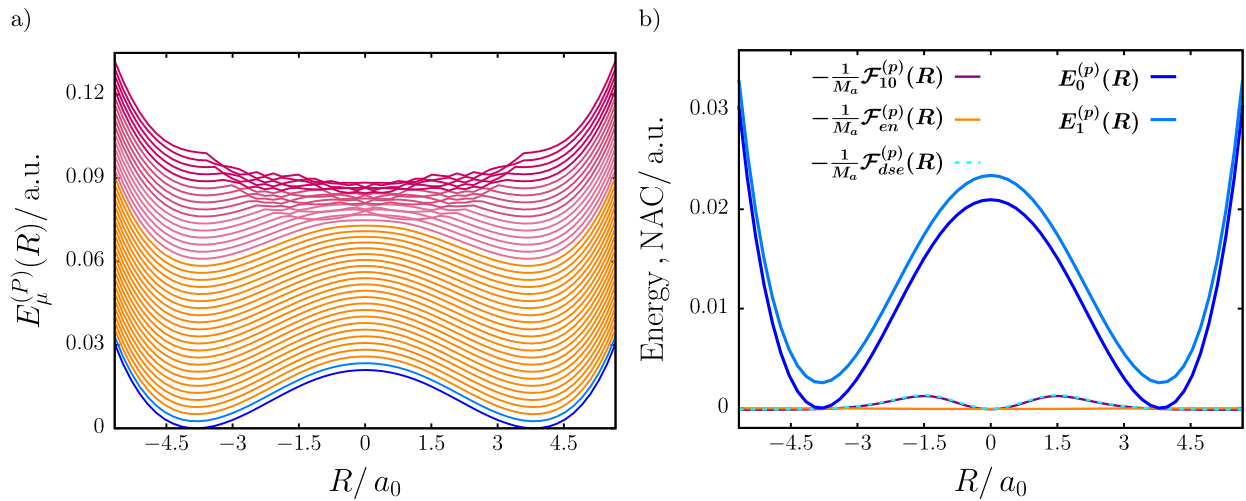}
\end{center}
\renewcommand{\baselinestretch}{1.}
\caption{a) 40 energetically lowest lying adiabatic polariton cPES, $E^{(p)}_\mu(R)$, as function of nuclear coordinate, $R$, obtained from, $\hat{H}_p$, for the CSM model in the weak NAC regime under vibrational strong coupling with $\eta=0.04$. Lowest two surfaces are coloured in blue and high energy cPES manifold subject to avoided crossings is coloured in red. b) Two energetically lowest lying polaritonic cPES, $E^{(p)}_0(R)$ and $E^{(p)}_1(R)$, besides corresponding mass-weighted nuclear derivative NAC element, $\mathcal{F}^{(p)}_{10}(R)$, with molecular, $\mathcal{F}^{(p)}_{\mathrm{en}}(R)$, and DSE contribution, $\mathcal{F}^{(p)}_{dse}(R)$, as function of nuclear coordinate, $R$.} 
\label{fig.vsc_polariton_length}
\end{figure*}

In Fig.\ref{fig.vsc_polariton_length}a, we show polaritonic cPESs, $E^{(p)}_\mu(R)$, obtained from $\hat{H}_p$ for forty states of the weakly non-adiabatic CSM model under VSC with $\eta=0.04$ (\textit{cf.} Appendix \ref{subsec.polariton_hamiltonian} for numerical details).
We find a dense manifold of surfaces (colored orange), which show an energetic separation of \textit{approximately} $\hbar\omega_c$. At elevated energies several avoided crossings appear (colored red), which can be traced back to the first excited adiabatic electronic state becoming energetically dominant. 

In Fig.\ref{fig.vsc_polariton_length}b, the two energetically lowest lying cPES (coloured blue) and the corresponding mass-weighted, nuclear derivative NAC element, $-\frac{1}{M_a}\mathcal{F}^{(p)}_{10}(R)$, are shown as functions of the nuclear coordinate. The NAC element is subject to two symmetric maxima at around $R\approx\pm1.5\,a_0$, where $-\frac{1}{M_a}\mathcal{F}^{(p)}_{\mathrm{en}}(R)=0.0012\,\mathrm{a.u.}$ and cPES are separated by $\Delta E_{10}=0.0024\,\mathrm{a.u.}=527\,\mathrm{cm}^{-1}$. From the latter, we have, $\Delta E_{10}<\hbar\omega_c\approx585\,\mathrm{cm}^{-1}$, which indicates strong light-matter coupling effects, since the first excited state's cPES is not just shifted by the cavity frequency relative to the ground state cPES. Further, the nuclear derivative NAC element shows a significant magnitude compared to the energy gap. We find, $\mathcal{F}^{(p)}_{10}$, to be dominantly determined by the cavity-induced contribution to the nuclear derivative NAC, $-\frac{1}{M_a}\mathcal{F}^{(p)}_{dse}(R)$ (\textit{cf.} Appendix \ref{subsec.polariton_hamiltonian}), as shown in dashed-blue, \textit{i.e.}, non-adiabatic coupling is an effect of strong interaction between electrons and cavity modes. The molecular component, $-\frac{1}{M_a}\mathcal{F}^{(p)}_{en}(R)$, is very small and not observable from Fig.\ref{fig.vsc_polariton_length}b. 

Based on the non-vanishing non-adiabatic coupling and the small energetic spacing of polaritonic cPES obained from $\hat{H}_p$, the CBO approximation for the ground state cPES seems not to be well justified from this perspective. Furthermore, due to small energetic separation of polaritonic cPES, corresponding ``vibrational'' eigenstates are assumed to hybridize, such that the bare ground state polaritonic cPES does not contain all relevant information. From a dynamics perspective, a ground state truncation additionally implies that the light-matter hybrid system will dynamically evolve only on a single surface, which effectively relates to suppression of excitations in the cavity subspace. Consequently, the \textit{electron-polariton} Hamiltonian, $\hat{H}_p$, seems not straightforwardly applicable to ground state problems under the CBO approximation as compared to VSC theory discussed in Sec.\ref{sec.vsc_theory}.

We like to close by emphasizing that our arguments apply to cavities, which operate at \textit{low energies} compared to electronic excitations, \textit{i.e.}, infrared or terahertz cavities with, $\hbar\omega_c\approx\Delta_\mathrm{vib}\ll\Delta_e$. In particular, our arguments \textit{do not} transfer to scenarios, where cavity mode frequencies acquire a significant fraction of electronic excitation energies as recently studied computationally in context of cavity-altered ground state reactions\cite{pavosevic2022a,pavosevic2022b,mandal2023}.

\section{Summary and Conclusion}
\label{sec.summary}
In this paper, we extended \textit{ab initio} vibro-polaritonic chemistry beyond the cavity Born-Oppenheimer approximation\cite{flick2017,flick2017cbo} and discussed the role of electron-photon correlation in the vibrational strong coupling regime. In the first part, we quantitatively reviewed VSC theory by derivation of coupled TISE for adiabatic electron-photon states and nuclear-photon states. Electrons are here described by an extended electronic structure problem, which accounts for correlations between electrons and cavity modes.\cite{flick2017cbo} We furthermore derived explicit expressions for nuclear and cavity non-adiabatic derivative coupling elements via the generalized Hellmann-Feynman theorem. In combination with a numerical analysis of non-adiabatic coupling in the cavity Shin-Metiu model under VSC, our findings can be formulated as:
\begin{enumerate}
	\item[1.] Cavity-induced non-adiabatic coupling is characterized by transition dipole moments between adiabatic electron-photon states and therefore relies crucially on their optical character. Nuclear non-adiabatic coupling is altered in presence of cavity modes by acquiring a transition dipole mediated correction to the vibronic coupling of adiabatic electron-photon states. The additional term emerges from the dipole self-energy and can be understood in terms of the transverse polarization operator.
	\item[2.] The CBO approximation is well justified, when both adiabatic electron-photon states are energetically well separated and light-matter non-adiabatic coupling is small. The latter relates in particular to vanishing transition dipole moments between electron-photon states. Numerically, we find cavity-induced non-adiabatic coupling relevant if the system is highly excited in the cavity mode, which can become relevant even at low energies, when the independent (uncoupled from cavity modes) molecular system is already subject to strong \textit{vibronic} coupling.
	\item[3.]  Numerical investigation concerning the applicability of the electron-polariton Hamiltonian in the VSC regime revealed the emergence of many close lying polaritonic cPES, which can be locally subject to cavity-induced non-adiabatic coupling. Thus, the CBO approximation seems not simply applicable in this setting. We argue that a separation of electrons as ``fast'' and cavity modes in combination with nuclei as ``slow'' degrees of freedom finds in VSC theory a more suitable theoretical approach to vibro-polaritonic chemistry. In contrast, $\hat{H}_p$, applies well in scenarios, where \textit{both} electrons \textit{and} cavity modes resemble ''fast'' high-energy degrees of freedom and nuclei are ``slowly'' moving, \textit{i.e.}, the electronic strong coupling regime.
\end{enumerate}

In the second part of this paper, we introduced a consistent approach to widely employed effective ground state models in vibro-polaritonic chemistry, named crude VSC theory, which can be understood as an approximation to the CBO model. Our formulation relies on a crude VSC Born-Huang expansion that employs a basis of adiabatic electronic states and \textit{does not} satisfy the CBO but a \textit{crude} CBO approximation. Accordingly, in crude VSC theory, only the bare electronic structure problem is solved \textit{ab initio}, which motivates our argument that crude CBO and CBO ground state theories differ in the way they account for electron-photon correlation. We identified correlations partially missing in the crude framework to stem from neglected non-adiabatic coupling terms related to transition dipole moments in the adiabatic subspace, which in turn leads to different electron-photon entanglement character in $\Psi_\mathrm{cbo}$ and $\Psi_\mathrm{ccbo}$. 

We furthermore proposed a perturbative formulation of CBO ground state theory inspired by molecular Herzberg-Teller theory, denoted as crude VSC perturbation theory (cVSC-PT). In the cVSC-PT framework, we characterized the \textit{crude} CBO ground state as first-order approximation to the CBO ground state, which accounts only for electron-photon entanglement arising indirectly from interactions of electrons and cavity modes with nuclei. Correlation corrections manifest then as cavity-induced, transition dipole-mediated non-adiabatic couplings between adiabatic electronic ground and excited states, which are shown to initially enter at second-order of cVSC-PT. Further characteristics of electron-photon correlation effects based on a numerical analysis of the CSM model summarize as follows:

\begin{enumerate}
	\item[4.] Electron-photon correlation in low-lying adiabatic electron-photon states manifests in terms of density variations in the molecular subspace relative to light-matter uncorrelated adiabatic electronic states (non-interacting limit of light-matter hybrid system).
	\item[5.] Classical activation barriers on CBO cPES are subject to electron-photon correlation effects, which are absent on crude cPES. In the CSM model, electron-photon correlation induces a barrier increase of $4\,\%$ relative to the crude cPES under VSC at $\eta=0.04$. Minimum energy paths on cPES are in contrast well reproduced by effective ground state models concerning path curvature, despite shortcomings in related reaction potentials. 
	\item[6.] Perturbative electron-photon correlation corrections at second- and third-order of cVSC-PT were explicitly evaluated for the weakly non-adiabatic CSM model under VSC. Non-vanishing corrections illustrated the relevance of electron-photon correlation beyond the crude CBO approximation. Further, cVSC-PT corrections allowed us to obtain significantly improved classical activation barriers relative to the crude cPES of the CSM model.
\end{enumerate}

We conclude by pointing out that the cVSC-PT Ansatz allows for estimating when an effective ground state description in the crude CBO approximation can qualitatively capture trends of the fully correlated CBO approach. Since \textit{ab initio} approaches to the extended electronic structure problem underlying fully correlated VSC theory are still in their infancy, it would be desirable to identify suitable scenarios, where one can benefit from the appealing nature of crude model approaches to vibro-polaritonic chemistry. 
Further, non-adiabatic effects under vibrational strong coupling will become relevant for reactive systems subject to energetically low-lying excited states. Such ``beyond CBO'' scenarios would be furthermore conceptually interesting, since they point at a deeper understanding of strong interactions between electrons, nuclei and cavity modes. 
Finally, collective effects in vibro-polaritonic chemistry and their inclusion in an \textit{ab initio} description of light-matter hybrid systems under VSC pose a contemporary theoretical issue relevant for connecting theoretical predictions with experimental findings. Crude VSC theory with perturbative electron-photon correlations corrections relative to fully correlated approaches can also here constitute a beneficial first step towards collective strong coupling scenarios due to its inherently simpler character.

\section*{Acknowledgements}
We acknowledge fruitful discussions with Tillmann Klamroth, Christoph Witzorky and Foudhil Bouakline (all Potsdam). This work was funded by the Deutsche Forschungsgemeinschaft (DFG, German Research Foundation) under Germany's Excellence Strategy - EXC 2008/1-390540038. E.W. Fischer also acknowledges support by the International Max Planck Research School for Elementary Processes in Physical Chemistry at the Fritz Haber Institute.

\section*{Data Availability Statement}
The data that support the findings of this study are available from the corresponding author upon reasonable request.

\section*{Conflict of Interest}
The authors have no conflicts to disclose.

\renewcommand{\thesection}{}
\section*{Appendix}

\setcounter{equation}{0}
\renewcommand{\theequation}{\thesubsection.\arabic{equation}}
\subsection{Details on Derivative NAC Elements}
\label{subsec.appendix_derivative_nacs}
We analytically evaluate derivatives of electron-photon Hamiltonians with respect to cavity displacement and nuclear coordinates in VSC theory for derivative NAC elements as discussed in Sec. \ref{subsec.length_nacs_details}. 
\subsubsection{Cavity Derivative NAC Elements}
For the cavity displacement coordinate derivative, we obtain 
\begin{align}
\dfrac{\partial}{\partial x_{\lambda k}}
\hat{H}_{ec}
&=
\dfrac{\partial}{\partial x_{\lambda k}}
\left(
\hat{H}_e
+
V_c
+
\hat{H}_{sc}
+
\hat{H}_{dse}
\right)
\quad,
\vspace{0.2cm}
\\
&=
\dfrac{\partial}{\partial x_{\lambda k}}
\sum^{2N_c}_{\lambda^\prime, k^\prime}
\biggl(
\dfrac{\omega^2_{k^\prime}}{2}
x^2_{\lambda^\prime k^\prime}
\vspace{0.2cm}
\\
&
\hspace{1cm}
+
\sqrt{\dfrac{2\omega_{k^\prime}}{\hbar}}\,
g_{k^\prime}\,
\biggl(
\underline{e}_{\lambda^\prime k^\prime}
\cdot
\underline{d}_{en}
\biggr)\,
x_{\lambda^\prime k^\prime}
\nonumber
\vspace{0.2cm}
\\
&
\hspace{2cm}
+
\dfrac{g^2_{k^\prime}}{\hbar\omega_{k^\prime}}
\biggl(
\underline{e}_{\lambda^\prime k^\prime}
\cdot
\underline{d}_{en}
\biggr)^2
\biggr)
\nonumber
\quad,
\vspace{0.2cm}
\\
&=
\omega^2_k\,
x_{\lambda k}
+
\sqrt{\dfrac{2\omega_k}{\hbar}}\,
g_k\,
\biggl(
\underline{e}_{\lambda k}
\cdot
\underline{d}_{en}
\biggr)
\quad,
\end{align}
such that the corresponding matrix element evaluates to
\begin{multline}
\braket{\Psi^{(ec)}_\nu
\vert
\dfrac{\partial}{\partial x_{\lambda k}}\,\hat{H}_{ec}
\vert
\Psi^{(ec)}_\mu}_{\underline{r}}
=
\omega^2_k\,
x_{\lambda k}\,
\delta_{\nu\mu}
\\
+
\sqrt{\dfrac{2\omega_k}{\hbar}}\,
g_k
\biggl(
\underline{e}_{\lambda k}
\cdot
\underline{\mathcal{D}}_{\nu\mu}
\biggr)
\quad.
\end{multline}
Here, the diagonal term in the first line vanishes since, $\nu\neq\mu$. 

\subsubsection{Nuclear Derivative NAC Elements}
For the nuclear coordinate derivative, we find
\begin{align}
\underline{\nabla}_a
\hat{H}_{ec}
&=
\underline{\nabla}_a
\hat{H}_{e}
+
\underline{\nabla}_a
V_c
+
\underline{\nabla}_a
\hat{H}_{sc}
+
\underline{\nabla}_a
\hat{H}_{dse}
\quad,
\end{align}
where we immediately recognize, $\underline{\nabla}_a V_c=\underline{0}$. The interaction term's derivative gives
\begin{align}
\underline{\nabla}_a
\hat{H}_{sc}
&=
\underline{\nabla}_a
\sum^{2N_c}_{\lambda,k}
\sqrt{\dfrac{2\omega_k}{\hbar}}\,
g_k\,
\biggl(
\underline{e}_{\lambda k}
\cdot
\underline{d}_{en}
\biggr)\,
x_{\lambda k}
\quad,
\vspace{0.2cm}
\\
&=
\sum^{2N_c}_{\lambda,k}
\sqrt{\dfrac{2\omega_k}{\hbar}}\,
g_k\,
\underline{\nabla}_a
\biggl(
\underline{e}_{\lambda k}
\cdot
\underline{d}_n
\biggr)
x_{\lambda k}
\quad,
\vspace{0.2cm}
\\
&=
\sum^{2N_c}_{\lambda,k}
\sqrt{\dfrac{2\omega_k}{\hbar}}\,
g_k\,
\underline{\nabla}_a
\sum_b
Q_b\,
\biggl(
\underline{e}_{\lambda k}
\cdot
\underline{R}_b
\biggr)
x_{\lambda k}
\,,
\end{align}
where in the second line, $\underline{\nabla}_a\biggl(\underline{e}_{\lambda k}\cdot\underline{d}_e\biggr)=\underline{0}$. With, $\underline{d}_n=\sum_b Q_b\,\underline{R}_b$, in the third line and evaluation of nuclear gradients as
\begin{align}
\underline{\nabla}_a
\biggl(
\underline{e}_{\lambda k}
\cdot
\underline{R}_b
\biggr)
&=
\underline{e}_{\lambda k}
\biggl(
\underline{\nabla}_a
\cdot
\underline{R}_b
\biggr)
=
\underline{e}_{\lambda k}\,
\delta_{ab}
\quad,
\end{align}
it follows that
\begin{align}
\underline{\nabla}_a
\hat{H}_{sc}
&=
\sum^{2N_c}_{\lambda,k}
\sqrt{\dfrac{2\omega_k}{\hbar}}\,
g_k\,
Q_a\,
\underline{e}_{\lambda k}\,
x_{\lambda k}
\quad.
\end{align}
Turning to the dipole self-energy term, the nuclear derivative follows as
\begin{align}
\underline{\nabla}_a
\hat{H}_{dse}
&=
\underline{\nabla}_a
\sum^{2N_c}_{\lambda, k}
\dfrac{g^2_k}{\hbar\omega_k}\,
\biggl(
\underline{e}_{\lambda k}
\cdot
\underline{d}_{en}
\biggr)^2
\quad,
\vspace{0.2cm}
\\
&=
\sum^{2N_c}_{\lambda, k}
\dfrac{2\,g^2_k}{\hbar\omega_k}\,
\biggl(
\underline{e}_{\lambda k}
\cdot
\underline{d}_{en}
\biggr)\,
\underline{\nabla}_a\,
\biggl(
\underline{e}_{\lambda k}
\cdot
\underline{d}_{en}
\biggr)
\,,
\vspace{0.2cm}
\\
&=
\sum^{2N_c}_{\lambda, k}
\dfrac{2\,g^2_k}{\hbar\omega_k}\,
\biggl(
\underline{e}_{\lambda k}
\cdot
\underline{d}_{en}
\biggr)\,
Q_a\,
\underline{e}_{\lambda k}
\quad,
\end{align}
where we employed the chain rule in the second line and took advantage of the result found for the interaction term in the third line. The nuclear derivative of the DSE term is related to the transverse polarization operator\cite{schaefer2020}  
\begin{align}
\underline{\hat{P}}_\perp
&=
\dfrac{1}{4\pi}
\sum^{2N_c}_{\alpha}
\biggl(
\underline{\lambda}_{\alpha}
\cdot
\underline{d}_{en}
\biggr)\,
\underline{\lambda}_{\alpha}
\quad,
\vspace{0.2cm}
\\
&=
\dfrac{1}{4\pi}
\sum^{2N_c}_{\lambda,k}
\lambda^2_k\,
\biggl(
\underline{e}_{\lambda k}
\cdot
\underline{d}_{en}
\biggr)\,
\underline{e}_{\lambda k}
\quad,
\vspace{0.2cm}
\\
&=
\dfrac{1}{4\pi}
\sum^{2N_c}_{\lambda,k}
\dfrac{2g^2_k}{\hbar\omega_k}\,
\biggl(
\underline{e}_{\lambda k}
\cdot
\underline{d}_{en}
\biggr)\,
\underline{e}_{\lambda k}
\quad,
\end{align}
where the second line follows with, $\alpha=(\lambda,k)$, for the summation index and, $\underline{\lambda}_{\lambda k}=\lambda_k\,\underline{e}_{\lambda k}$, whereas in the third line, $\lambda_k=\sqrt{\frac{2}{\hbar\omega_k}}\,g_k$. For the corresponding matrix element, we find
\begin{multline}
\braket{\Psi^{(ec)}_\nu
\vert
\underline{\nabla}_a\,\hat{H}_{ec}
\vert
\Psi^{(ec)}_\mu}_{\underline{r}}
=
\braket{\Psi^{(ec)}_\nu
\vert
\underline{\nabla}_a
\hat{H}_{e}
\vert
\Psi^{(ec)}_\mu}_{\underline{r}}
\\
+
\braket{\Psi^{(ec)}_\nu
\vert
\underline{\nabla}_a
\hat{H}_{sc}
\vert
\Psi^{(ec)}_\mu}_{\underline{r}}
\\
+
\braket{\Psi^{(ec)}_\nu
\vert
\underline{\nabla}_a\,\hat{H}_{dse}
\vert
\Psi^{(ec)}_\mu}_{\underline{r}}
\quad,
\end{multline}
where the second term corresponding to the light-matter interaction contribution, $\underline{\nabla}_a\hat{H}_{sc}$, is independent of electronic coordinates and vanishes therefore due to orthogonality of adiabatic electronic states. 

\setcounter{equation}{0}
\renewcommand{\theequation}{\thesubsection.\arabic{equation}}
\subsection{Details on Crude VSC Theory}
\label{subsec.crude_vsc_details}
Starting from Eq.\eqref{eq.crude_projected_molecule_photon_tise_length}, the matrix element with respect to $\hat{H}_{ec}$ is evaluated as
\begin{align}
\bra{\Psi^{(e)}_{\nu}}
\hat{H}_{ec}\,
\tilde{\chi}^{(nc)}_{i\mu}\,
\ket{\Psi^{(e)}_{\mu}}
&=
\left(
E^{(e)}_\nu(\underline{R})
+
V_c(\underline{x})
\right)
\tilde{\chi}^{(nc)}_{i\nu}
\label{eq.crude_projected_electron_photon_hamiltonian}
\vspace{0.2cm}
\\&
+
\braket{\Psi^{(e)}_{\nu}
\vert
\hat{H}_{sc}
\vert
\Psi^{(e)}_{\mu}}_{\underline{r}}\,
\tilde{\chi}^{(nc)}_{i\mu}
\nonumber
\vspace{0.2cm}
\\&
+
\braket{\Psi^{(e)}_{\nu}
\vert
\hat{H}_{dse}
\vert
\Psi^{(e)}_{\mu}}_{\underline{r}}\,
\tilde{\chi}^{(nc)}_{i\mu}
\nonumber
\quad,
\end{align}
where the first term constitutes the diagonal contribution determined by the molecular PES, $E^{(e)}_\nu(\underline{R})$, and the cavity potential, $V_c(\underline{x})$. The second term resembles the matrix element for the light-matter interaction, $\hat{H}_{sc}$, which evaluates to
\begin{multline}
\bra{\Psi^{(e)}_{\nu}}
\hat{H}_{sc}
\ket{\Psi^{(e)}_{\mu}}_{\underline{r}}\,
\tilde{\chi}^{(nc)}_{i\mu}
\\
=
\sum^{2N_c}_{\lambda,k}
\sqrt{\dfrac{2\omega_k}{\hbar}}\,
g_k
\biggl(
\underline{e}_{\lambda k}
\cdot
\underline{d}_{\nu\mu}
\biggr)\,
x_{\lambda k}\,
\tilde{\chi}^{{(nc)}}_{i\mu}
\quad,
\end{multline}
and has in general both diagonal and off-diagonal contributions with dipole matrix elements as given in Eq.\eqref{eq.crude_dipole_matrix_elements}. The third term in Eq.\eqref{eq.crude_projected_electron_photon_hamiltonian} involves the dipole self-energy term, $\hat{H}_{dse}$, with matrix element 
\begin{multline}
\bra{\Psi^{(e)}_{\nu}}
\hat{H}_{dse}\,
\tilde{\chi}^{(nc)}_{i\mu}\,
\ket{\Psi^{(e)}_{\mu}}
\\
=
\sum^{2N_c}_{\lambda,k}
\dfrac{g^2_k}{\hbar\omega_k}
\sum_\alpha
\biggl(
\underline{e}_{\lambda k}
\cdot
\underline{d}_{\nu\alpha}
\biggr)
\biggl(
\underline{e}_{\lambda k}
\cdot
\underline{d}_{\alpha\mu}
\biggr)
\tilde{\chi}^{{(nc)}}_{i\mu}
\quad,
\end{multline}
where we inserted the resolution-of-the-identity in the adiabatic subspace, $\sum_\alpha\ket{\Psi^{(e)}_\alpha}\bra{\Psi^{(e)}_\alpha}=\boldsymbol{1}^{(e)}$, between projected dipole moments. For diagonal matrix elements with $\nu=\mu$, it is instructive to rewrite the sum over adiabatic states as
\begin{multline}
\sum_{\alpha}
\biggl(
\underline{e}_{\lambda k}
\cdot
\underline{d}_{\nu\alpha}
\biggr)
\biggl(
\underline{e}_{\lambda k}
\cdot
\underline{d}_{\alpha\nu}
\biggr)
=
\sum_{\alpha}
\biggl(
\underline{e}_{\lambda k}
\cdot
\underline{d}^2_{\nu\alpha}
\biggr)
\quad,
\end{multline}
which holds, when $\underline{e}_{\lambda k}$ projects out only a single element of the molecular transition dipole moment, $\underline{d}_{\nu\alpha}$. This can be always achieved for linearly polarized cavity modes as considered here: Vectors, $\{\underline{e}_{\lambda k},\underline{e}_{\lambda^\prime k}, \underline{k}\}$, span a cavity coordinate frame, which can be chosen parallel to the molecule fixed frame with molecular center of mass located at the origin. Then, only single elements of $\underline{d}_{\nu\alpha}$ are addressed with respect to polarization vectors, $\underline{e}_{\lambda k}$, and, $\underline{e}_{\lambda^\prime k}$, parallel to the molecular frame's axis.

\setcounter{equation}{0}
\renewcommand{\theequation}{\thesubsection.\arabic{equation}}
\subsection{Details on Reduced Density Matrices}
\label{subsec.rdm_details}
A beneficial connection between CBO and crude CBO ground states in Eqs.\eqref{eq.vsc_cbo_ground_state} and \eqref{eq.vsc_crude_cbo_ground_state} can be introduced by realizing that the adiabatic electron-photon ground state, $\ket{\Psi^{(ec)}_0(\underline{R},\underline{x})}$, at a fixed nuclear configuration, $\underline{R}$, can be expanded in an orthonormal basis of adiabatic electronic states, $\ket{\Psi^{(e)}_\mu(\underline{R})}$.\cite{izmaylov2017} Accordingly, the CBO ground state can be written as
\begin{align}
\ket{\Psi_\mathrm{cbo}(\underline{R},\underline{x})}
&=
\chi^{(nc)}_{0}(\underline{R},\underline{x})
\left(
\sum_\mu
c_\mu(\underline{x})\,
\ket{\Psi^{(e)}_\mu(\underline{R})}
\right)
\vspace{0.2cm}
\\
&=
\sum_\mu
\tilde{\chi}^{(nc)}_\mu(\underline{R},\underline{x})\,
\ket{\Psi^{(e)}_\mu(\underline{R})}
\quad,
\label{eq.cbo_state_adiabatic_expansion}
\end{align}
where we absorb expansion coefficients, $c_\mu(\underline{x})$, which account for the cavity displacement coordinate dependence, in the second line into, $\tilde{\chi}^{(nc)}_\mu(\underline{R},\underline{x})=c_\mu(\underline{x})\,\chi^{(nc)}_{0}(\underline{R},\underline{x})$. Eq.\eqref{eq.cbo_state_adiabatic_expansion} is formally appealing, since the \textit{crude} CBO ground state, $\ket{\Psi_\mathrm{ccbo}(\underline{R},\underline{x})}$, introduced in Eq.\eqref{eq.vsc_crude_cbo_ground_state} follows directly by truncating Eq.\eqref{eq.cbo_state_adiabatic_expansion} after the first term. 

Further, one can similarly expand the crude CBO ground state in an orthonormal basis of crude adiabatic \textit{electronic} states, $\ket{\Psi^{(e)}_i}=\ket{\Psi^{(e)}_i(\underline{R}_0)}$, with nuclear reference configuration, $\underline{R}_0$, leading to
\begin{align}
\ket{\Psi_\mathrm{ccbo}(\underline{R},\underline{x})}
&=
\tilde{\chi}^{(nc)}_{0}(\underline{R},\underline{x})
\left(
\sum_i
u_i(\underline{R})\,
\ket{\Psi^{(e)}_i}
\right)
\vspace{0.2cm}
\\
&=
\sum_i
\tilde{\varphi}^{(nc)}_i(\underline{R},\underline{x})\,
\ket{\Psi^{(e)}_i}
\quad,
\label{eq.crude_cbo_state_adiabatic_expansion}
\end{align}
where expansion coefficients, $u_i(\underline{R})$, now accounting for nuclear coordinate dependence, are absorbed into, $\tilde{\varphi}^{(nc)}_i(\underline{R},\underline{x})=u_i(\underline{R})\,\tilde{\chi}^{(nc)}_{0}(\underline{R},\underline{x})$. By truncating the expansion in Eq.\eqref{eq.crude_cbo_state_adiabatic_expansion} after the first term, we obtain a zeroth-order reference (ref) state
\begin{align}
\ket{\Psi_\mathrm{ref}(\underline{R},\underline{x})}
&=
\tilde{\varphi}^{(nc)}_0(\underline{R},\underline{x})\,
\ket{\Psi^{(e)}_0}
\quad,
\label{eq.crude_reference_state}
\end{align}
which is a product state with respect to electronic and nuclear-cavity degrees of freedom, \textit{i.e.}, electrons are by definition disentangled from nuclei and cavity modes. 

In the following, electronic, nuclear and cavity reduced density matrices are derived for the CBO ground state, the crude CBO ground state and the partially disentangled reference state in Eqs.\eqref{eq.cbo_state_adiabatic_expansion}-\eqref{eq.crude_reference_state}.

\subsubsection{Electronic Reduced Density Matrices}
Starting with electronic RDM, we obtain for the CBO ground state 
\begin{align}
\hat{\rho}^{(e)}_\mathrm{cbo}(\underline{r},\underline{r}^\prime)
&=
\displaystyle\iint
\mathrm{d}\underline{R}\,
\mathrm{d}\underline{x}\,
\Psi_\mathrm{cbo}(\underline{r},\underline{R},\underline{x})\,
\Psi^\star_\mathrm{cbo}(\underline{r}^\prime,\underline{R},\underline{x})
\quad,
\nonumber
\vspace{0.2cm}
\\
&=
\sum_{\mu\nu}
\displaystyle\int\mathrm{d}\underline{R}\,
\Psi^{(e)}_\mu(\underline{r};\underline{R})\,
\Psi^{(e)\star}_\nu(\underline{r}^\prime;\underline{R})
\nonumber
\\
&\hspace{1.2cm}
\times
\displaystyle\int\mathrm{d}\underline{x}\,
\tilde{\chi}_\mu(\underline{R},\underline{x})\,
\tilde{\chi}^\star_\nu(\underline{R},\underline{x})
\quad,
\nonumber
\vspace{0.2cm}
\\
&=
\sum_{\mu\nu}
\displaystyle\int\mathrm{d}\underline{R}\,
\rho_{\mu\nu}(\underline{r},\underline{r}^\prime;\underline{R})\,
\tilde{S}_{\mu\nu}(\underline{R},\underline{R})
\quad,
\nonumber
\vspace{0.2cm}
\\
&=
\sum_{\mu\nu}
\underbrace{\tilde{\rho}_{\mu\nu}(\underline{r},\underline{r}^\prime)}_{=\tilde{\rho}^{(e)}_{\mu\nu}}
\quad,
\label{eq.electron_rdm_cbo}
\end{align}
with vibro-polaritonic overlap integral
\begin{align}
\tilde{S}_{\mu\nu}(\underline{R},\underline{R})
&=
\displaystyle\int\mathrm{d}\underline{x}\,
\tilde{\chi}_\mu(\underline{R},\underline{x})\,
\tilde{\chi}^\star_\nu(\underline{R},\underline{x})
=
\tilde{S}^{(n)}_{\mu\nu}
\quad,
\end{align}
and
\begin{align}
\tilde{\rho}_{\mu\nu}(\underline{r},\underline{r}^\prime)\,
&=
\displaystyle\int\mathrm{d}\underline{R}\,
\rho_{\mu\nu}(\underline{r},\underline{r}^\prime;\underline{R})\,
\tilde{S}_{\mu\nu}(\underline{R},\underline{R})
=
\tilde{\rho}^{(e)}_{\mu\nu}
\,,
\end{align}
where
\begin{align}
\rho_{\mu\nu}(\underline{r},\underline{r}^\prime;\underline{R})
&=
\Psi^{(e)}_\mu(\underline{r};\underline{R})\,
\Psi^{(e)\star}_\nu(\underline{r}^\prime;\underline{R})
\quad.
\end{align}
The crude CBO electronic RDM follows immediately by truncating the sum over adiabatic electronic states in Eq.\eqref{eq.electron_rdm_cbo} as
\begin{align}
\hat{\rho}^{(e)}_\mathrm{ccbo}(\underline{r},\underline{r}^\prime)
=
\tilde{\rho}^{(e)}_{00}
\quad.
\end{align}
For the reference state, one finds
\begin{align}
\hat{\rho}^{(e)}_\mathrm{ref}(\underline{r},\underline{r}^\prime)
&=
\displaystyle\iint
\mathrm{d}\underline{R}\,
\mathrm{d}\underline{x}\,
\Psi_\mathrm{ref}(\underline{r},\underline{R},\underline{x})\,
\Psi^\star_\mathrm{ref}(\underline{r}^\prime,\underline{R},\underline{x})
\quad,
\nonumber
\vspace{0.2cm}
\\
&=
\Psi^{(e)}_0(\underline{r})\,
\Psi^{(e)\star}_0(\underline{r}^\prime)
\underbrace{\displaystyle\iint
\mathrm{d}\underline{R}\,
\mathrm{d}\underline{x}\,
\vert
\tilde{\varphi}_0(\underline{R},\underline{x})
\vert^2}_{=1}
\quad,
\nonumber
\vspace{0.2cm}
\\
&=
\underbrace{
\Psi^{(e)}_0(\underline{r})\,
\Psi^{(e)\star}_0(\underline{r}^\prime)}_{=\rho^{(e)}_{00}}
\quad,
\end{align}
where normalization of $\tilde{\varphi}_0(\underline{R},\underline{x})$ was employed in the second line. Note, the third line resembles a bare (crude adiabatic) electronic contribution.

\subsubsection{Nuclear Reduced Density Matrices}
The nuclear RDM for the CBO state follows as
\begin{align}
\hat{\rho}^{(n)}_\mathrm{cbo}(\underline{R},\underline{R}^\prime)
&=
\displaystyle\iint
\mathrm{d}\underline{r}\,
\mathrm{d}\underline{x}\,
\Psi_\mathrm{cbo}(\underline{r},\underline{R},\underline{x})\,
\Psi^\star_\mathrm{cbo}(\underline{r},\underline{R}^\prime,\underline{x})
\quad,
\nonumber
\vspace{0.2cm}
\\
&=
\sum_{\mu\nu}
\displaystyle\int
\mathrm{d}\underline{x}\,
\tilde{\chi}_\mu(\underline{R},\underline{x})\,
\tilde{\chi}^\star_\nu(\underline{R}^\prime,\underline{x})
\nonumber
\\
&\hspace{1.2cm}
\times
\displaystyle\int
\mathrm{d}\underline{r}\,
\Psi^{(e)}_\mu(\underline{r};\underline{R})\,
\Psi^{(e)\star}_\nu(\underline{r};\underline{R}^\prime)
\quad,
\nonumber
\vspace{0.2cm}
\\
&=
\sum_{\mu\nu}
\underbrace{
\tilde{S}_{\mu\nu}(\underline{R},\underline{R}^\prime)\,
O_{\mu\nu}(\underline{R},\underline{R}^\prime)}_{=\tilde{S}^{(n)}_{\mu\nu}\,
O^{(n)}_{\mu\nu}}
\quad,
\label{eq.nuclear_rdm_cbo}
\end{align}
with electronic overlap integral
\begin{align}
O_{\mu\nu}(\underline{R},\underline{R}^\prime)
&=
\displaystyle\int\mathrm{d}\underline{r}\,
\Psi^{(e)}_\mu(\underline{r};\underline{R})\,
\Psi^{(e)\star}_\nu(\underline{r};\underline{R}^\prime)
=
O^{(n)}_{\mu\nu}
\quad.
\end{align}
Note, the adiabatic matrix element, $O_{\mu\nu}(\underline{R},\underline{R}^\prime)$, is non-zero for $\underline{R}\neq\underline{R}^\prime$. The crude CBO equivalent follows again by truncating the sum over adiabatic electronic states in Eq.\eqref{eq.nuclear_rdm_cbo} to the ground state contribution 
\begin{align}
\hat{\rho}^{(n)}_\mathrm{ccbo}(\underline{R},\underline{R}^\prime)
&=
\tilde{S}^{(n)}_{00}\,
O^{(n)}_{00}
\quad.
\end{align}
Eventually, for the reference state, we have
\begin{align}
\hat{\rho}^{(n)}_\mathrm{ref}(\underline{R},\underline{R}^\prime)
&=
\displaystyle\iint
\mathrm{d}\underline{r}\,
\mathrm{d}\underline{x}\,
\Psi_\mathrm{ref}(\underline{r},\underline{R},\underline{x})\,
\Psi^\star_\mathrm{ref}(\underline{r},\underline{R}^\prime,\underline{x})
\quad,
\nonumber
\vspace{0.2cm}
\\
&=
\displaystyle\int\mathrm{d}\underline{x}\,
\tilde{\varphi}_0(\underline{R},\underline{x})\,
\tilde{\varphi}^\star_0(\underline{R}^\prime,\underline{x})
\underbrace{
\displaystyle\int\mathrm{d}\underline{r}\,
\vert
\Psi^{(e)}_0(\underline{r})
\vert^2}_{=1}
\quad,
\nonumber
\vspace{0.2cm}
\\
&=
\underbrace{\tilde{s}_{00}(\underline{R},\underline{R}^\prime)}_{=\tilde{s}^{(n)}_{00}}
\quad,
\end{align}
where normalization of $\psi_0(\underline{r})$ was employed in the second line. Note, in contrast to CBO and crude CBO nuclear RDM, $\hat{\rho}^{(n)}_\mathrm{ref}$ is independent of adiabatic electronic contributions, which directly follows from the product nature of the reference state. 

\subsubsection{Cavity Reduced Density Matrices}
Finally, the cavity RDM for the CBO ground state is obtained as
\begin{align}
\hat{\rho}^{(c)}_\mathrm{cbo}(\underline{x},\underline{x}^\prime)
&=
\displaystyle\iint
\mathrm{d}\underline{r}\,
\mathrm{d}\underline{R}\,
\Psi_\mathrm{cbo}(\underline{r},\underline{R},\underline{x})\,
\Psi^\star_\mathrm{cbo}(\underline{r},\underline{R},\underline{x}^\prime)
\quad,
\nonumber
\vspace{0.2cm}
\\
&=
\sum_{\mu\nu}
\displaystyle\int\mathrm{d}\underline{R}\,
\tilde{\chi}_\mu(\underline{R},\underline{x})\,
\tilde{\chi}^\star_\nu(\underline{R},\underline{x}^\prime)
\nonumber
\\
&\hspace{1.2cm}
\times
\underbrace{
\displaystyle\int\mathrm{d}\underline{r}\,
\Psi^{(e)}_\mu(\underline{r};\underline{R})\,
\Psi^{(e)\star}_\nu(\underline{r};\underline{R})}_{=\delta_{\nu\mu}}
\quad,
\nonumber
\vspace{0.2cm}
\\
&=
\sum_\mu
\displaystyle\int\mathrm{d}\underline{R}\,
\tilde{\chi}_\mu(\underline{R},\underline{x})\,
\tilde{\chi}^\star_\mu(\underline{R},\underline{x}^\prime)
\quad,
\nonumber
\vspace{0.2cm}
\\
&=
\sum_\mu
\underbrace{\tilde{S}_{\mu\mu}(\underline{x},\underline{x}^\prime)}_{=\tilde{S}^{(c)}_{\mu\mu}}
\quad,
\end{align}
and the crude CBO cavity RDM is straightforwardly found to be
\begin{align}
\hat{\rho}^{(c)}_\mathrm{ccbo}(\underline{x},\underline{x}^\prime)
&=
\tilde{S}^{(c)}_{00}
\quad.
\end{align}
For the reference state, we find an expression similar to the nuclear RDM with
\begin{align}
\hat{\rho}^{(n)}_\mathrm{ref}(\underline{x},\underline{x}^\prime)
&=
\displaystyle\iint
\mathrm{d}\underline{r}\,
\mathrm{d}\underline{R}\,
\Psi_\mathrm{ref}(\underline{r},\underline{R},\underline{x})\,
\Psi^\star_\mathrm{ref}(\underline{r},\underline{R},\underline{x}^\prime)
\quad,
\nonumber
\vspace{0.2cm}
\\
&=
\displaystyle\int\mathrm{d}\underline{R}\,
\tilde{\varphi}_0(\underline{R},\underline{x})\,
\tilde{\varphi}^\star_0(\underline{R},\underline{x}^\prime)
\underbrace{
\displaystyle\int\mathrm{d}\underline{r}\,
\vert
\Psi^{(e)}_0(\underline{r})
\vert^2}_{=1}
\quad,
\nonumber
\vspace{0.2cm}
\\
&=
\underbrace{\tilde{s}_{00}(\underline{x},\underline{x}^\prime)}_{=\tilde{s}^{(c)}_{00}}
\quad.
\end{align}
As in the nuclear case, the latter is independent of adiabatic electronic contributions.

\setcounter{equation}{0}
\renewcommand{\theequation}{\thesubsection.\arabic{equation}}
\subsection{Numerical Details on the VSC-CSM Model}
\label{subsec.numerics_csm_model}
For the numerical solution of the length-gauge CSM model Hamiltonian, Eq.\eqref{eq.length_csm_hamiltonian}, we employ a Colbert-Miller discrete variable representation (DVR) for all degrees of freedom, \textit{i.e.}, the electron, the moving nucleus and the cavity mode, with KEO matrix elements
\begin{align}
T_{ij}
=
\dfrac{\hbar^2}{2\Delta s^2}\,
(-1)^{i-j}
\begin{cases}
\dfrac{\pi^2}{3},& i=j
\vspace{0.2em}
\\
\dfrac{2}{(i-j)^2}, & i\neq j
\end{cases},
\end{align}
for coordinates, $s=r,R,x_c$. Up to vibrational strong coupling with $\eta=0.04$, we obtain converged results on grids $r\in[-2L,2L]$, $R\in[-\frac{L}{2},+\frac{L}{2}]$, and, $x_c\in[-300,300]\,\sqrt{m_e}\,a_0$. We employ grid points, $M_e=M_n=M_c=201$, for electrons, nucleus and cavity mode in both the weak and strong non-adiabatic coupling regimes.

\setcounter{equation}{0}
\renewcommand{\theequation}{\thesubsection.\arabic{equation}}
\subsection{Details on cVSC-PT(3)}
\label{subsec.third_order_cpes_correction}
We provide an explicit expression for the third-order energy correction in cVSC-PT(3) when applied to the CSM model. Following conventional Rayleigh-Schr\"odinger perturbation theory, the third-order energy correction is given by 
\begin{align}
E^{(3)}_0
&=
\sum_{\mu,\nu\neq0}
\dfrac{
\Delta V_{0\mu}\,
\Delta V_{\mu\nu}\,
\Delta V_{\nu 0}
}
{\Delta E^{(e)}_{0\mu}\Delta E^{(e)}_{0\nu}}
-
\Delta V_{00}
\sum_{\mu\neq0}
\dfrac{
\vert
\Delta V_{0\mu}
\vert^2
}{\left(\Delta E^{(e)}_{0\mu}\right)^2}
\quad,
\label{eq.third_order_cpes_correction}
\end{align}
where we suppress coordinate dependence for brevity. We first show, that contributions involving the cavity potential, $V_c$, cancel. By taking into account the orthonormal character of adiabatic electronic states, the first term can be written as
\begin{multline}
\dfrac{\Delta V_{0\mu}\,
\Delta V_{\mu\nu}\,
\Delta V_{\nu 0}}
{\Delta E^{(e)}_{0\mu}\Delta E^{(e)}_{0\nu}}
=
\dfrac{V_c
\left(
H^{sc}_{0\mu}
+
H^{dse}_{0\mu}
\right)
\left(
H^{sc}_{\nu 0}
+
H^{dse}_{\nu 0}
\right)
\delta_{\mu\nu}}
{\Delta E^{(e)}_{0\mu}\Delta E^{(e)}_{0\nu}}
\\
+
\dfrac{\left(
H^{sc}_{0\mu}
+
H^{dse}_{0\mu}
\right)
\left(
H^{sc}_{\mu\nu}
+
H^{dse}_{\mu\nu}
\right)
\left(
H^{sc}_{\nu 0}
+
H^{dse}_{\nu 0}
\right)}{\Delta E^{(e)}_{0\mu}\Delta E^{(e)}_{0\nu}}
\quad,
\end{multline}
where the first contribution is diagonal in adiabatic state indices $\mu,\nu$, such that it cancels with the first term in
\begin{multline}
\dfrac{\Delta V_{00}\,
\vert
\Delta V_{0\mu}
\vert^2}
{\left(\Delta E^{(e)}_{0\mu}\right)^2}
=
\dfrac{V_c
\left(
H^{sc}_{0\mu}
+
H^{dse}_{0\mu}
\right)
\left(
H^{sc}_{\mu 0}
+
H^{dse}_{\mu 0}
\right)}
{\left(\Delta E^{(e)}_{0\mu}\right)^2}
\\
+
\dfrac{\left(
H^{sc}_{00}
+
H^{dse}_{00}
\right)
\left(
H^{sc}_{0\mu}
+
H^{dse}_{0\mu}
\right)
\left(
H^{sc}_{\mu 0}
+
H^{dse}_{\mu 0}
\right)}
{\left(\Delta E^{(e)}_{0\mu}\right)^2}
\quad,
\end{multline}
due to different signs in Eq.\eqref{eq.third_order_cpes_correction}. Thus, $E^{(3)}_0$ is independent of $V_c$ and can be decomposed into eight contributions, $E^{(3)}_0=\sum^8_{i=1}E^{(3)}_{0,i}$, by grouping equivalent products of expectation values. The leading order term scales as $g^3$ and reads
\small{
\begin{widetext}
\begin{align}
E^{(3)}_{0,1}
&=
\sum_{\mu,\nu\neq0}
\dfrac{
\Delta H^{sc}_{0\mu}\,
\Delta H^{sc}_{\mu\nu}\,
\Delta H^{sc}_{\nu 0}
}
{\Delta E^{(e)}_{0\mu}\Delta E^{(e)}_{0\nu}}
-
\Delta H^{sc}_{00}
\sum_{\mu\neq0}
\dfrac{
\vert
\Delta H^{sc}_{0\mu}
\vert^2
}{\left(\Delta E^{(e)}_{0\mu}\right)^2}
=
\sum_{\mu,\nu\neq0}
\dfrac{
(2\omega_c)^{\frac{3}{2}}\,
g^3\,x^3_c\,
d_{0\mu}\,
d_{\mu\nu}\,
d_{\nu 0}
}
{\Delta E^{(e)}_{0\mu}\Delta E^{(e)}_{0\nu}}
-
\sum_{\mu\neq0}
\dfrac{
(2\omega_c)^{\frac{3}{2}}\,
g^3\,x^3_c\,
d_{00}\,
d^2_{0\mu}
}{\left(\Delta E^{(e)}_{0\mu}\right)^2}
\quad,
\end{align}
\end{widetext}}
\normalsize
Terms with two interaction and a single dipole self-energy expectation value scale quartic in the light-matter interaction constant, $g^4$, and read explicitly
\small{
\begin{widetext}
\begin{align}
E^{(3)}_{0,2}
&=
\sum_{\mu,\nu\neq0}
\dfrac{
\Delta H^{sc}_{0\mu}\,
\Delta H^{sc}_{\mu\nu}\,
\Delta H^{dse}_{\nu 0}
}
{\Delta E^{(e)}_{0\mu}\Delta E^{(e)}_{0\nu}}
-
\Delta H^{sc}_{00}
\sum_{\mu\neq0}
\dfrac{
\Delta H^{sc}_{0\mu}
\Delta H^{dse}_{\mu 0}
}{\left(\Delta E^{(e)}_{0\mu}\right)^2}
=
\sum_{\mu,\nu\neq0}
\sum_{\alpha}
\dfrac{
2\,g^4\,x^2_c\,
d_{0\mu}\,
d_{\mu\nu}
\left(
d_{\nu\alpha}
d_{\alpha 0}
\right)}
{\Delta E^{(e)}_{0\mu}\Delta E^{(e)}_{0\nu}}
-
\sum_{\mu\neq0}
\sum_{\alpha}
\dfrac{
2\,g^4\,x^2_c\,
d_{00}\,
d_{0\mu}
\left(
d_{\mu\alpha}
d_{\alpha 0}
\right)
}{\left(\Delta E^{(e)}_{0\mu}\right)^2}
\,,
\vspace{0.2cm}
\\
E^{(3)}_{0,3}
&=
\sum_{\mu,\nu\neq0}
\dfrac{
\Delta H^{sc}_{0\mu}\,
\Delta H^{dse}_{\mu\nu}\,
\Delta H^{sc}_{\nu 0}
}
{\Delta E^{(e)}_{0\mu}\Delta E^{(e)}_{0\nu}}
-
\Delta H^{sc}_{00}
\sum_{\mu\neq0}
\dfrac{
\Delta H^{dse}_{0\mu}
\Delta H^{sc}_{\mu 0}
}{\left(\Delta E^{(e)}_{0\mu}\right)^2}
=
\sum_{\mu,\nu\neq0}
\sum_\alpha
\dfrac{
2\,g^4\,x^2_c\,
d_{0\mu}\,
\left(
d_{\mu\alpha}
d_{\alpha\nu}
\right)
d_{\nu 0}}
{\Delta E^{(e)}_{0\mu}\Delta E^{(e)}_{0\nu}}
-
\sum_{\mu\neq0}
\sum_\alpha
\dfrac{
2\,g^4\,x^2_c\,
d_{00}\,
\left(
d_{0\alpha}
d_{\alpha\mu}
\right)
d_{\mu0}
}{\left(\Delta E^{(e)}_{0\mu}\right)^2}
\,,
\vspace{0.2cm}
\\
E^{(3)}_{0,4}
&=
\sum_{\mu,\nu\neq0}
\dfrac{
\Delta H^{dse}_{0\mu}\,
\Delta H^{sc}_{\mu\nu}\,
\Delta H^{sc}_{\nu 0}
}
{\Delta E^{(e)}_{0\mu}\Delta E^{(e)}_{0\nu}}
-
\Delta H^{dse}_{00}
\sum_{\mu\neq0}
\dfrac{
\vert
\Delta H^{sc}_{0\mu}
\vert^2
}{\left(\Delta E^{(e)}_{0\mu}\right)^2}
=
\sum_{\mu,\nu\neq0}
\sum_\alpha
\dfrac{
2\,g^4\,x^2_c\,
\left(
d_{0\alpha}
d_{\alpha\mu}
\right)
d_{\mu\nu}\,
d_{\nu 0}}
{\Delta E^{(e)}_{0\mu}\Delta E^{(e)}_{0\nu}}
-
\sum_{\mu\neq0}
\sum_\alpha
\dfrac{
2\,g^4\,x^2_c\,
\left(
d_{0\alpha}
d_{\alpha 0}
\right)
d^2_{0\mu}
}{\left(\Delta E^{(e)}_{0\mu}\right)^2}
\,.
\end{align}
\end{widetext}}
\normalsize
The quintic contribution scaling as $g^5$ is determined by three terms
\small{
\begin{widetext}
\begin{align}
E^{(3)}_{0,5}
&=
\sum_{\mu,\nu\neq0}
\dfrac{
\Delta H^{sc}_{0\mu}\,
\Delta H^{dse}_{\mu\nu}\,
\Delta H^{dse}_{\nu 0}
}
{\Delta E^{(e)}_{0\mu}\Delta E^{(e)}_{0\nu}}
-
\Delta H^{sc}_{00}
\sum_{\mu\neq0}
\dfrac{
\Delta H^{dse}_{0\mu}
\Delta H^{dse}_{\mu 0}
}{\left(\Delta E^{(e)}_{0\mu}\right)^2}
\quad,
\nonumber
\vspace{0.2cm}
\\
&=
\sum_{\mu,\nu\neq0}
\sum_{\alpha,\beta}
\sqrt{\dfrac{2}{\omega^3_c}}
\dfrac{g^5\,x_c\,
d_{0\mu}
\left(
d_{\mu\alpha}
d_{\alpha\nu}
\right)
\left(
d_{\nu\beta}
d_{\beta 0}
\right)}
{\Delta E^{(e)}_{0\mu}\Delta E^{(e)}_{0\nu}}
-
\sum_{\mu\neq0}
\sum_{\alpha,\beta}
\sqrt{\dfrac{2}{\omega^3_c}}
\dfrac{g^5\,x_c\,
d_{00}
\left(
d_{0\alpha}
d_{\alpha\mu}
\right)
\left(
d_{\mu\beta}
d_{\beta 0}
\right)
}{\left(\Delta E^{(e)}_{0\mu}\right)^2}
\quad,
\vspace{0.2cm}
\\
E^{(3)}_{0,6}
&=
\sum_{\mu,\nu\neq0}
\dfrac{
\Delta H^{dse}_{0\mu}\,
\Delta H^{sc}_{\mu\nu}\,
\Delta H^{dse}_{\nu 0}
}
{\Delta E^{(e)}_{0\mu}\Delta E^{(e)}_{0\nu}}
-
\Delta H^{dse}_{00}
\sum_{\mu\neq0}
\dfrac{
\Delta H^{sc}_{0\mu}
\Delta H^{dse}_{\mu 0}
}{\left(\Delta E^{(e)}_{0\mu}\right)^2}
\quad,
\nonumber
\vspace{0.2cm}
\\
&=
\sum_{\mu,\nu\neq0}
\sum_{\alpha,\beta}
\sqrt{\dfrac{2}{\omega^3_c}}
\dfrac{g^5\,x_c\,
\left(
d_{0\alpha}
d_{\alpha\mu}
\right)
d_{\mu\nu}
\left(
d_{\nu\beta}
d_{\beta 0}
\right)}
{\Delta E^{(e)}_{0\mu}\Delta E^{(e)}_{0\nu}}
-
\sum_{\mu\neq0}
\sum_{\alpha,\beta}
\sqrt{\dfrac{2}{\omega^3_c}}
\dfrac{g^5\,x_c\,
\left(
d_{0\alpha}
d_{\alpha0}
\right)
d_{0\mu}
\left(
d_{\mu\beta}
d_{\beta 0}
\right)
}{\left(\Delta E^{(e)}_{0\mu}\right)^2}
\quad,
\vspace{0.2cm}
\\
E^{(3)}_{0,7}
&=
\sum_{\mu,\nu\neq0}
\dfrac{
\Delta H^{dse}_{0\mu}\,
\Delta H^{dse}_{\mu\nu}\,
\Delta H^{sc}_{\nu 0}
}
{\Delta E^{(e)}_{0\mu}\Delta E^{(e)}_{0\nu}}
-
\Delta H^{dse}_{00}
\sum_{\mu\neq0}
\dfrac{
\Delta H^{dse}_{0\mu}
\Delta H^{sc}_{\mu 0}
}{\left(\Delta E^{(e)}_{0\mu}\right)^2}
\nonumber
\vspace{0.2cm}
\\
&=
\sum_{\mu,\nu\neq0}
\sum_{\alpha,\beta}
\sqrt{\dfrac{2}{\omega^3_c}}
\dfrac{g^5\,x_c\,
\left(
d_{0\alpha}
d_{\alpha\mu}
\right)
\left(
d_{\mu\beta}
d_{\beta\nu}
\right)
d_{\nu 0}}
{\Delta E^{(e)}_{0\mu}\Delta E^{(e)}_{0\nu}}
-
\sum_{\mu\neq0}
\sum_{\alpha,\beta}
\sqrt{\dfrac{2}{\omega^3_c}}
\dfrac{g^5\,x_c\,
\left(
d_{0\alpha}
d_{\alpha0}
\right)
\left(
d_{0\beta}
d_{\beta\mu}
\right)
d_{\mu 0}
}{\left(\Delta E^{(e)}_{0\mu}\right)^2}
\quad,
\end{align}
\end{widetext}}
\normalsize
and contain a single interaction and two DSE expectation values. Eventually, the remaining term scales as $g^6$ and contains three DSE expectation values
\small{
\begin{widetext}
\begin{align}
E^{(3)}_{0,8}
&=
\sum_{\mu,\nu\neq0}
\dfrac{
\Delta H^{dse}_{0\mu}\,
\Delta H^{dse}_{\mu\nu}\,
\Delta H^{dse}_{\nu 0}
}
{\Delta E^{(e)}_{0\mu}\Delta E^{(e)}_{0\nu}}
-
\Delta H^{dse}_{00}
\sum_{\mu\neq0}
\dfrac{
\Delta H^{dse}_{0\mu}
\Delta H^{dse}_{\mu 0}
}{\left(\Delta E^{(e)}_{0\mu}\right)^2}
\quad,
\nonumber
\vspace{0.2cm}
\\
&=
\sum_{\mu,\nu\neq0}
\sum_{\alpha,\beta,\gamma}
\dfrac{g^6}{\omega^3_c}
\dfrac{
\left(
d_{0\alpha}
d_{\alpha\mu}
\right)
\left(
d_{\mu\beta}
d_{\beta\nu}
\right)
\left(
d_{\nu\gamma}
d_{\gamma 0}
\right)}
{\Delta E^{(e)}_{0\mu}\Delta E^{(e)}_{0\nu}}
-
\sum_{\mu\neq0}
\sum_{\alpha,\beta,\gamma}
\dfrac{g^6}{\omega^3_c}
\dfrac{
\left(
d_{0\alpha}
d_{\alpha0}
\right)
\left(
d_{0\beta}
d_{\beta\mu}
\right)
\left(
d_{\mu\gamma}
d_{\gamma 0}
\right)
}{\left(\Delta E^{(e)}_{0\mu}\right)^2}
\quad.
\end{align}
\end{widetext}}
\normalsize

\setcounter{equation}{0}
\renewcommand{\theequation}{\thesubsection.\arabic{equation}}
\subsection{Details on the Electron-Polariton Hamiltonian}
\label{subsec.polariton_hamiltonian}
\subsubsection{Extended Born-Huang Expansion and TISEs}
The electron-polariton Hamiltonian is given as
\begin{align}
\hat{H}_p
&=
\hat{T}_c
+
\hat{H}_{ec}
\quad,
\label{eq.electron_polariton_pauli_fierz}
\end{align}
and augments the length-gauge electron-photon Hamiltonian by the cavity KEO. A generalized Born-Huang expansion can be introduced as
\begin{align}
\Psi_i(\underline{r},\underline{R},\underline{x})
&=
\sum_\mu
\phi^{(n)}_{i\mu}(\underline{R})\,
\Psi^{(p)}_\mu(\underline{r},\underline{x};\underline{R})
\quad,
\label{eq.electron_polariton_born_huang}
\end{align} 
with orthonormal, adiabatic \textit{electron-polariton} states, $\Psi^{(p)}_\mu(\underline{r},\underline{x};\underline{R})$, and nuclear states, $\phi^{(n)}_{i\mu}(\underline{R})$, providing expansion coefficients. Adiabatic states treat electrons and cavity modes here on equal footing and only the nuclear coordinate dependence is understood to be parametric in nature. The molecular Pauli-Fierz Hamiltonian takes the form, $\hat{H}_\mathrm{PF}=\hat{T}_n+\hat{H}_p$. 

After projecting on adiabatic electron-polariton states, $\bra{\Psi^{(p)}_\nu(\underline{R})}$, we obtain a system of two coupled TISEs, where adiabatic states solve
\begin{align}
\hat{H}_p
\ket{\Psi^{(p)}_{\nu}(\underline{R})}
&=
E^{(p)}_\nu(\underline{R})
\ket{\Psi^{(p)}_{\nu}(\underline{R})}
\quad,
\label{eq.electron_polariton_tise}
\end{align}
with polaritonic cPES, $E^{(p)}_\nu(\underline{R})$, which only depends on nuclear coordinates, $\underline{R}$, and provides a potential for a nuclear TISE
\begin{multline}
\left(
\hat{T}_n
+
E^{(p)}_\nu(\underline{R})
\right)
\phi^{(n)}_{i\nu}(\underline{R})
+
\sum_{\mu\neq\nu}
\mathcal{\hat{C}}^{(p)}_{\nu\mu}\,
\phi^{(n)}_{i\mu}(\underline{R})
\\
=
\mathcal{E}_i\,
\phi^{(n)}_{i\nu}(\underline{R})
\quad.
\label{eq.electron_polariton_nuclear_photon_tise}
\end{multline}
Non-adiabatic coupling elements, $\mathcal{\hat{C}}^{(p)}_{\nu\mu}$, are formally similar to Eq.\eqref{eq.nuclear_derivative_nac}, however, act now in the subspace of adiabatic \textit{electron-polariton} states. Nuclear derivative NAC elements take the form (\textit{cf.} Sec.\ref{subsec.length_nacs_details})
\begin{align}
\underline{\mathcal{F}}^{(p)}_{a,\nu\mu}(\underline{R})
&=
\dfrac{\braket{\Psi^{(p)}_\nu
\vert
\underline{\nabla}_a
V_{en}
\vert
\Psi^{(p)}_\mu}_{\underline{r},\underline{x}}}
{E^{(p)}_\nu-E^{(p)}_\mu}
\vspace{0.2cm}
\\
&
+
\displaystyle\sum^{2N_c}_{\lambda, k}
\dfrac{2\,Q_a\,g^2_k}{\hbar\omega_k}
\dfrac{
\biggl(
\underline{e}_{\lambda k}
\cdot
\mathcal{\underline{D}}^{(p)}_{\nu\mu}
\biggr)\,
\underline{e}_{\lambda k}}
{E^{(p)}_\nu-E^{(p)}_\mu}
\quad,
\nonumber
\end{align}
where the second line relates to a contribution of the transverse polarization operator (\textit{cf.} Appendix \ref{subsec.appendix_derivative_nacs}) in the basis of adiabatic electron-polariton states with transition dipole moments 
\begin{align}
\mathcal{\underline{D}}^{(p)}_{\nu\mu}(\underline{R})
&=
\braket{\Psi^{(p)}_\nu
\vert
\underline{d}_{en}
\vert
\Psi^{(p)}_\mu}_{\underline{r},\underline{x}}
\quad.
\end{align}
Although, $\underline{\mathcal{F}}^{(p)}_{a,\nu\mu}(\underline{R})$, is formally similar to $\underline{\mathcal{F}}^{(n)}_{a,\nu\mu}(\underline{R})$ in Eq.\eqref{eq.length_nuclear_nac_hellmann_feynman} (note different superscripts $(p)$ and $(n)$), it differs in the following aspect: For infrared Fabry-P\'erot cavities and ground/first excited states relevant for the CBO approximation, it holds that
\begin{align}
\vert E^{(p)}_1-E^{(p)}_0\vert
\approx
\hbar\omega_c
\ll
\Delta_e
\approx
\vert E^{(ec)}_1-E^{(ec)}_0\vert
\quad,
\end{align}	
with characteristic fundamental electronic transition energy, $\Delta_e$. On the left-hand side, the energy difference relates to polaritonic cPESs, $E^{(p)}_\nu$, obtained from TISE \eqref{eq.electron_polariton_tise}. In contrast, the energy difference on the right-hand side results from cPESs of VSC theory, $E^{(ec)}_\nu$, in Eq.\eqref{eq.length_vsc_electron_photon_tise}. For the latter, we assume two energetically well separated states, such that the CBO approximation is valid. We discuss properties of $\underline{\mathcal{F}}^{(p)}_{a,\nu\mu}(\underline{R})$ in some detail in Subsec.\ref{subsec.elec_pol_gs}.

\subsubsection{Numerical Details}
We solve the electron-polariton TISE \eqref{eq.electron_polariton_tise} illustratively for the CSM model in the weak NAC regime with $R_c=1.5\,\text{\AA}$ and $\eta=0.04$. Convergence of the $40$ energetically lowest lying polaritonic cPESs is obtained with grids, $r\in[-2L,2L]$, $x_c\in[-300,\,300]\,\sqrt{m_e}\,a_0$ and $R\in[-\frac{L}{2},\,+\frac{L}{2}]$, discretized by grid points, $M_e=91$, $M_c=91$ and $M_n=101$. Note, we restrict the electronic grid to the region where the electron-nuclear potential dominates and do not account for regions far from nuclei as determined by a harmonically-confining electronic DSE contribution. The latter may be relevant if the electron is driven far from the three nuclei due to an external driving field for example, which is beyond the scope of the present work.



\begin{thebibliography}{200}
\bibitem{cavity_mode_summation} We write, $\displaystyle\sum^2_\lambda\displaystyle\sum^{N_c}_k=\displaystyle\sum^{2N_c}_{\lambda,k}$, for summation over polarization and cavity mode indices.
\end{thebibliography}
\end{document}